\documentclass[fleqn,usenatbib]{mnras}

\usepackage{newtxtext,newtxmath}

\usepackage[T1]{fontenc}
\usepackage{ae,aecompl}

\usepackage{graphicx}	
\usepackage{amsmath}	
\graphicspath{{./plots/}}
\usepackage{xcolor}

\newcommand{\msun}{M$_{\odot}$}

\newcommand{\kms}{km~s$^{-1}$}
\newcommand{\ergs}{erg s$^{-1}$}
\newcommand{\Ha}{H$\alpha$}

\newcommand{\HeI}{He~{\sc i}}

\newcommand{\OI}{O~{\sc i}}
\newcommand{\OII}{O~{\sc ii}}

\newcommand{\Oneb}{[O~{\sc i}] \lam\lam6300, 6364}
\newcommand{\CII}{C~{\sc ii}}
\newcommand{\NaI}{Na~{\sc i}}
\newcommand{\MgII}{Mg~{\sc ii}}
\newcommand{\MgI}{Mg~{\sc i}}

\newcommand{\SiII}{Si~{\sc ii}}

\newcommand{\CaII}{Ca~{\sc ii}}
\newcommand{\caiif}{[Ca~{\sc ii}] \lam\lam7291, 7323}

\newcommand{\FeII}{Fe~{\sc ii}}
\newcommand{\FeIII}{Fe~{\sc iii}}

\newcommand{\Fefs}{$^{56}$Fe}
\newcommand{\Cofs}{$^{56}$Co}
\newcommand{\Nifs}{$^{56}$Ni}
\newcommand{\mej}{$M_\mathrm{ej}$}
\newcommand{\ek}{$E_\mathrm{k}$}
\newcommand{\vph}{$v_\mathrm{ph}$}

\newcommand{\eom}{$E_\mathrm{k}/M_\mathrm{ej}$}
\newcommand{\lam}{$\lambda$}

\newcommand{\Eh}{$E\left(B-V\right)_\mathrm{host}$}
\newcommand{\Emw}{$E\left(B-V\right)_\mathrm{MW}$}

\newcommand{\mni}{$M_\mathrm{Ni}$}

\newcommand{\sn}{SN\,}

\newcommand{\mzams}{$M_\mathrm{ZAMS}$}

\title[Luminous transitional SNe]{ Transitional events in the spectrophotometric regime between stripped envelope and superluminous supernovae}

\author[S. J. Prentice et al.]{S.~J.~Prentice,$^{1}$\thanks{E-mail: sipren.astro@gmail.com}, C.~Inserra$^{2}$, S.~Schulze$^{3}$, M.~Nicholl$^{4}$, P.~A.~Mazzali$^{5,6}$,  
\newauthor
S.~D.~Vergani$^{7}$, L. Galbany$^{8,9}$, J.~P.~Anderson$^{10}$, C.~Ashall$^{11}$, T.~W.~Chen$^{3}$,   
\newauthor
M.~Deckers$^{1}$, M.~Delgado~Manche\~no$^{8}$, R.~Gonz\'alez~D\'iaz$^{9,12}$, S.~ Gonz\'alez-Gait\'an$^{13}$, 
\newauthor
M.~Gromadzki$^{14}$, C.~P.~Guti\'errez$^{15,16}$, L.~Harvey$^{1}$, A.~Kozyreva$^{6}$, M.~R.~Magee$^{1}$,
\newauthor
 K.~Maguire$^{1}$,  T.~E.~M\"uller-Bravo$^{17}$, S.~Mu\~noz~Torres$^{8}$,  P.~J.~Pessi$^{10,18}$, J.~Sollerman$^{3}$,  
\newauthor
J.~Teffs$^{5}$, J.~H.~Terwel$^{1}$, and D.~R.~Young$^{19}$, 
\\
$^{1}$School of Physics, Trinity College Dublin, The University of Dublin, Dublin 2, Ireland\\
$^{2}$School of Physics \& Astronomy, Cardiff University, Queens Buildings, The Parade, Cardiff, CF24 3AA, UK\\
$^{3}$The Oskar Klein Centre, Department of Astronomy, Stockholm University, AlbaNova, SE-106 91 Stockholm , Sweden\\
$^{4}$Birmingham Institute for Gravitational Wave Astronomy and School of Physics and Astronomy, University of Birmingham, Birmingham B15 2TT, UK\\
$^{5}$Astrophysics Research Institute, Liverpool John Moores University, IC2, Liverpool Science Park, 146 Brownlow Hill,   Liverpool L3 5RF, UK\\
$^{6}$Max-Planck-Institut f{\"u}r Astrophysik, Karl-Schwarzschild-Str. 1, D-85748 Garching, Germany\\
$^{7}$GEPI, Observatoire de Paris, PSL University, CNRS, 5 Place Jules Janssen, 92190 Meudon, France\\
$^{8}$Departamento de F\'isica Te\'orica y del Cosmos, Universidad de Granada, E-18071 Granada, Spain\\
$^{9}$Institute of Space Sciences (ICE, CSIC), Campus UAB, Carrer de Can Magrans, s/n, E-08193 Barcelona, Spain\\
$^{10}$European Southern Observatory, Alonso de C\'ordova 3107, Casilla 19, Santiago, Chile\\
$^{11}$Institute for Astronomy, University of Hawai’i at Manoa, 2680 Woodlawn Dr., Hawai’i, HI 96822, USA\\
$^{12}$Instituto Nacional de Astrof{\'i}sica, {\'O}ptica y Electr{\'o}nica (INAOE), 72840 Tonantzintla, Puebla, Mexico\\
$^{13}$CENTRA-Centro de Astrof\'{\i}sica e Gravita\c{c}\~ao and Departamento de F\'{\i}sica, Instituto Superior T\'ecnico, \\Universidade de Lisboa, Avenida Rovisco Pais, 1049-001 Lisboa, Portugal\\
$^{14}$Astronomical Observatory, University of Warsaw, Al. Ujazdowskie 4, 00-478 Warszawa, Poland\\
$^{15}$Finnish Centre for Astronomy with ESO (FINCA), FI-20014 University of Turku, Finland\\
$^{16}$Tuorla Observatory, Department of Physics and Astronomy, FI-20014 University of Turku, Finland\\
$^{17}$School of Physics and Astronomy, University of Southampton, Southampton, Hampshire, SO17 1BJ, UK\\
$^{18}$Facultad de Ciencias Astronómicas y Geofísicas (FCAG), Universidad Nacional de La Plata (UNLP), Paseo del bosque S/N, 1900, Argentina\\
$^{19}$Astrophysics Research Centre, School of Mathematics and Physics, Queen's University Belfast, Belfast BT7 1NN, UK \\
}

\date{Accepted XXX. Received YYY; in original form ZZZ}

\pubyear{2021}

\begin{document}
\label{firstpage}
\pagerange{\pageref{firstpage}--\pageref{lastpage}}
\maketitle

\begin{abstract}
 The division between stripped-envelope supernovae (SE-SNe) and superluminous supernovae (SLSNe) is not well defined in either photometric or spectroscopic space. While a sharp luminosity threshold has been suggested, there remains an increasing number of transitional objects that reach this threshold without the spectroscopic signatures common to SLSNe. In this work we present data and analysis on four SNe transitional between SE-SNe and SLSNe; the He-poor SNe 2019dwa and 2019cri, and the He-rich SNe 2019hge and 2019unb. 
Each object displays long-lived and variable photometric evolution with luminosities around the SLSN threshold of $M_r < -19.8$ mag. Spectroscopically however, these objects are similar to SE-SNe, with line velocities lower than either SE-SNe and SLSNe, and thus represent an interesting case of rare transitional events.
\end{abstract}




\section{Introduction}

The classification scheme of stripped-envelope supernovae (SE-SNe) is a spectroscopic one that mostly evolved over the latter half of the Twentieth Century \citep[e.g.,][]{Filippenko1997}.
Type I and Type II separate H-poor and H-rich transients. Type Ia separates thermonuclear events from the Type Ibc SE-SNe. The latter group separated into the He-rich Type Ib and the He-poor Type Ic. Some SE-SNe, of Type Ic in particular, display broad and blended absorption features in their spectra indicative of a high specific kinetic energy. These are then further separated into Type Ic-BL.
To complicate matters, SE-SNe also include stars that explode with residual H in their outer layers, these are Type IIb.
\citet{Prentice2017} went further, separating the Type Ic into subclasses which expanded on the absorption width theme by counting the number of certain features in the spectra. This led to a sequence Ic-7 to Ic-3 with increasing specific kinetic energy, thus linking the taxonomic scheme with physical parameters. 

Since the discovery of hydrogen-poor superluminous supernovae (SLSNe) in the last decade or so \citep{Quimby2011}, an active topic of research has been establishing how these objects are connected to normal SE-SNe, including SNe Ib, Ic, IIb, broad line SNe, and supernova associated with gamma-ray bursts \citep[For studies of SE-SNe see][]{Lyman2016,Prentice2016,Prentice2019}.
`Superluminous supernova' however, is a classification mainly based upon a photometric property (luminosity).
Initially, SLSNe were separated in luminosity space from SE-SNe by an empirical cut: SLSNe are found at $M_r= -20 $ to $-22$ mag, making them amongst the most luminous transients in the universe \citep{GalYam2012}. 
Stripped-envelope supernovae are found at lower luminosities compared to SLSNe, $M_r = -16$ to $-19$ mag \citep[e.g.,][]{Drout2011,Taddia2018}.
With larger and more homogeneous samples, this initially useful magnitude boundary has become blurry: \citet{Angus2019} compared the luminosity distributions of SLSNe from the Dark Energy Survey (DES) with the literature sample of SLSNe, and found that the DES SLSN luminosity distribution peaks at $M_{4000} \sim -19.5$ mag, while the literature sample peaks at $M_{4000} \sim -20.75$ mag \citep[$M_{4000}$ is an artificial bandpass centred on 4000 \AA, see][]{inserra2014}.
\citet{DeCia2018} showed that if one assumes that SE-SNe and SLSNe are transients of similar origin then the luminosity distribution of these objects is smooth and decreases for increasing luminosity.

To have a phenomenological
definition of SLSNe similar to those of other SN types, the spectra evolution up to 30 days post-maximum needs to be probed. A SLSN I spectrum at 30 days resembles that of a type Ic
at peak \citep[e.g.][]{pastorello2010,inserra2013}, exhibiting a photospheric velocity that does not evolve
after 30+ days in contrast with typical SE-SNe \citep[e.g.][]{nicholl2015,liu2017,inserra2018}.

Another issue is that SLSNe are rare \citep{quimby2013,frohmaier2021}, and being luminous they are observed at higher redshift than other SN types making them relatively dim in the observer frame. Consequently, spectroscopic observations with good S/N are hard to come by, and long-term monitoring less likely. Thus the spectroscopic properties of the existing SLSN sample are not well sampled, with the exception of a few objects \citep[e.g. SNe 2015bn and 2017egm,][]{Nicholl2016,Bose2018}.

\citet{inserra2018} provided a statistical analysis of SLSNe-I identifying two subclasses based on their photometric and spectroscopic evolution together with the ejecta velocity.
\citet{Quimby2018} analysed a large sample of Palomar Transient Factory (PTF) SLSN-I and reported a similar finding with most objects following a similar spectroscopic evolution to PTF12dam or \sn2011ke. 
They also note that a few objects were `SLSN-like' but did not fit within this system or had limited data. One of these objects is PTF10ghi (henceforth \sn2010md) which showed H/He in its spectra, making it more SN IIb-like \citep{inserra2013,Quimby2018}.
Indeed, despite the obvious lack of He in the spectra of most objects \citet{Mazzali2016} demonstrated through spectral models that He may be present in the post-peak spectra in some SLSN-I. They identify iPTF13ajg as a likely candidate and suggested reclassifying this as a SN Ib. Although the photospheric-phase properties of SLSNe are clearly quite heterogeneous, a clustering analysis of their nebular phase spectra by \citet{Nicholl2019} could not identify multiple populations, suggesting their interiors may be similar despite differences in their envelopes and environments.

Another possible distinction between SLSNe (and some energetic SN~Ic) and SE-SNe is the observation that they are hosted in different type of galaxies. SLSNe occur in galaxies of lower metallicity and higher specific star-formation rate than is typical for SE-SNe \citep[e.g.,][]{Wiseman2020,Schulze2020}. One possibility is that the SLSN rate is suppressed in galaxies above a metallicity threshold of around one-half solar \citep{Perley2016,Chen2017a,Schulze2020}. The high specific star-formation rate has been used to point to SLSNe having progenitors with larger \mzams\ \citep{Leloudas2015}.

It is then clear that the division between SLSNe and normal SE-SNe is not as simple as originally proposed. In recent years, wide-field surveys have revealed a few unusual transients including transitional objects, in terms of luminosity or spectra evolution (or both) between SLSNe and normal CC-SNe \citep{Modjaz2019} such as SN~2017ens \citep{Chen2018}, a transition between a SLSN and a SN~IIn or SN~2019hcc (Parrag et al. submitted), a SN~II with a classification spectrum typical of a SLSN.  In this work, we present data and analysis on four photometrically unusual transients that sit in the `spectro-luminosity' region between SLSNe and SE-SNe, but which are spectroscopically similar to normal SE-SNe rather than to most SLSNe. Two of these objects, SNe 2019hge and 2019unb, are spectroscopically similar to both normal He-rich SE-SNe and to \sn2010md, but with lower luminosity.
They have previously been the subject of a study as part of a Zwicky Transient Facility \citep[ZTF;][]{Bellm2019} sample of SLSNe \citep{Yan2020}.

\begin{itemize}
    \item \sn2019cri/ZTF19aanijpu/ATLAS19gnt/Gaia19cpo/PS20axc was discovered by ZTF  on 2019-03-25 07:03:21. It was classified as a Type Ic-7 on 2019-04-23 08:47:39 \citep{2019TNSCR.625....1P}.

    \item \sn2019dwa/ZTF19aarfyvc/Gaia19bxj was discovered by ZTF on 2019-04-10 07:42:31. It was classified as a Type Ic supernova on 2019-05-16 20:05:46 \citep{2019TNSCR.799....1F}.

    \item \sn2019hge/ZTF19aawfbtgATLAS19och/Gaia19est/PS19elv was another ZTF discovery from 2019-05-31 11:11:30. It was classified as a peculiar SN IIb on 2019-08-23 11:25:41 \citep{2019TNSAN..74....1P}, and later as a SLSN-I \citep{2019TNSCR2859....1D}.

    \item Finally, \sn2019unb/ZTF19acgjpgh/ATLAS19bari/Gaia19fbu
    /PS19isr was discovered in the ZTF public stream and reported to the Transient Name Server\footnote{https://www.wis-tns.org/} (TNS) by the Trinity College Dublin transient group on 2019-10-20 12:28:30 \citep{2019TNSTR2335....1P}. It was classified as a peculiar Type II \citep{2019TNSCR2339....1P} and later a SLSN-I \citep{2020TNSCR1504....1D}.
\end{itemize}

\section{Data collection and reduction}

\begin{figure*}
    \centering
    \includegraphics[scale=0.45]{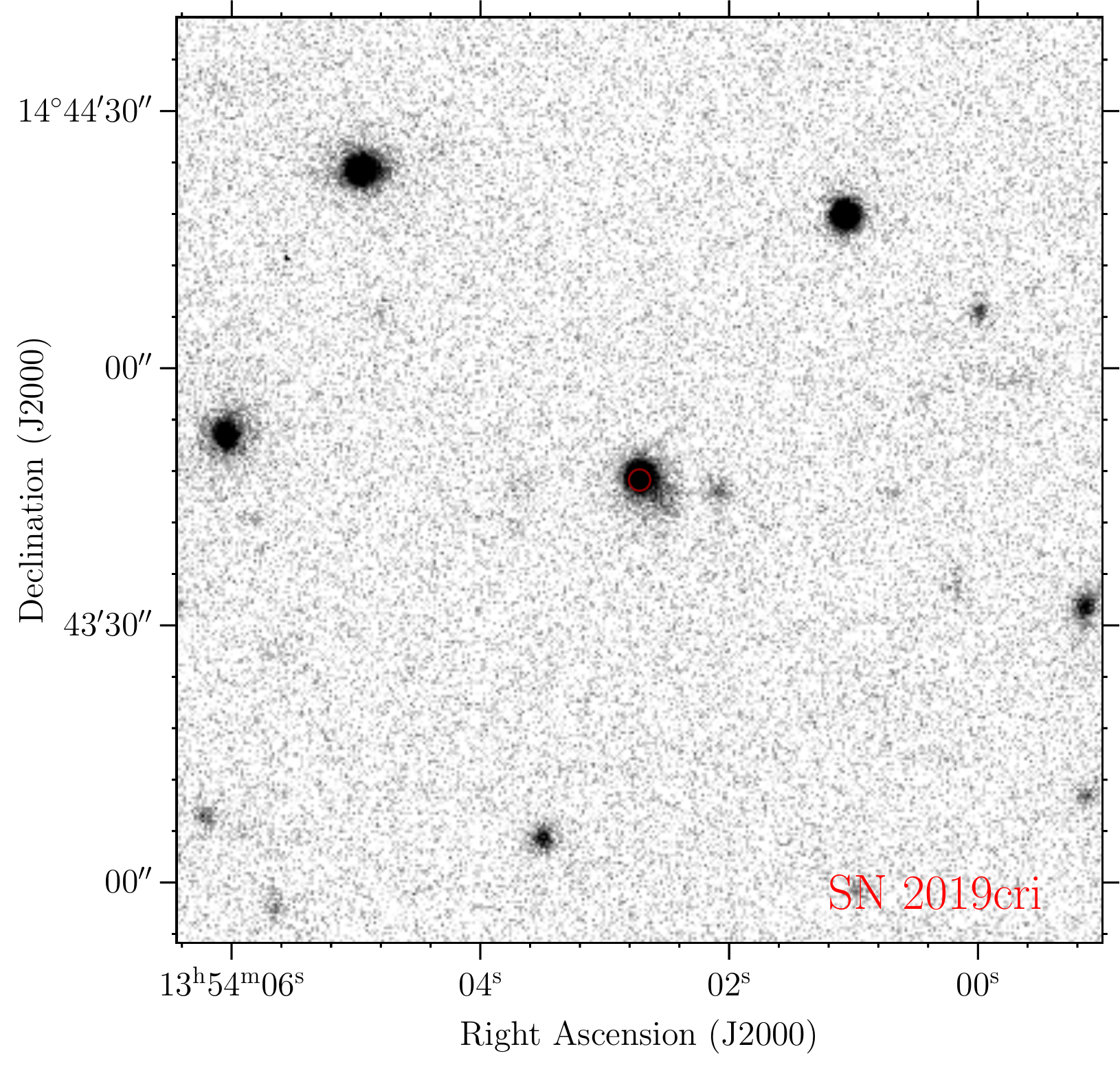}
    \includegraphics[scale=0.45]{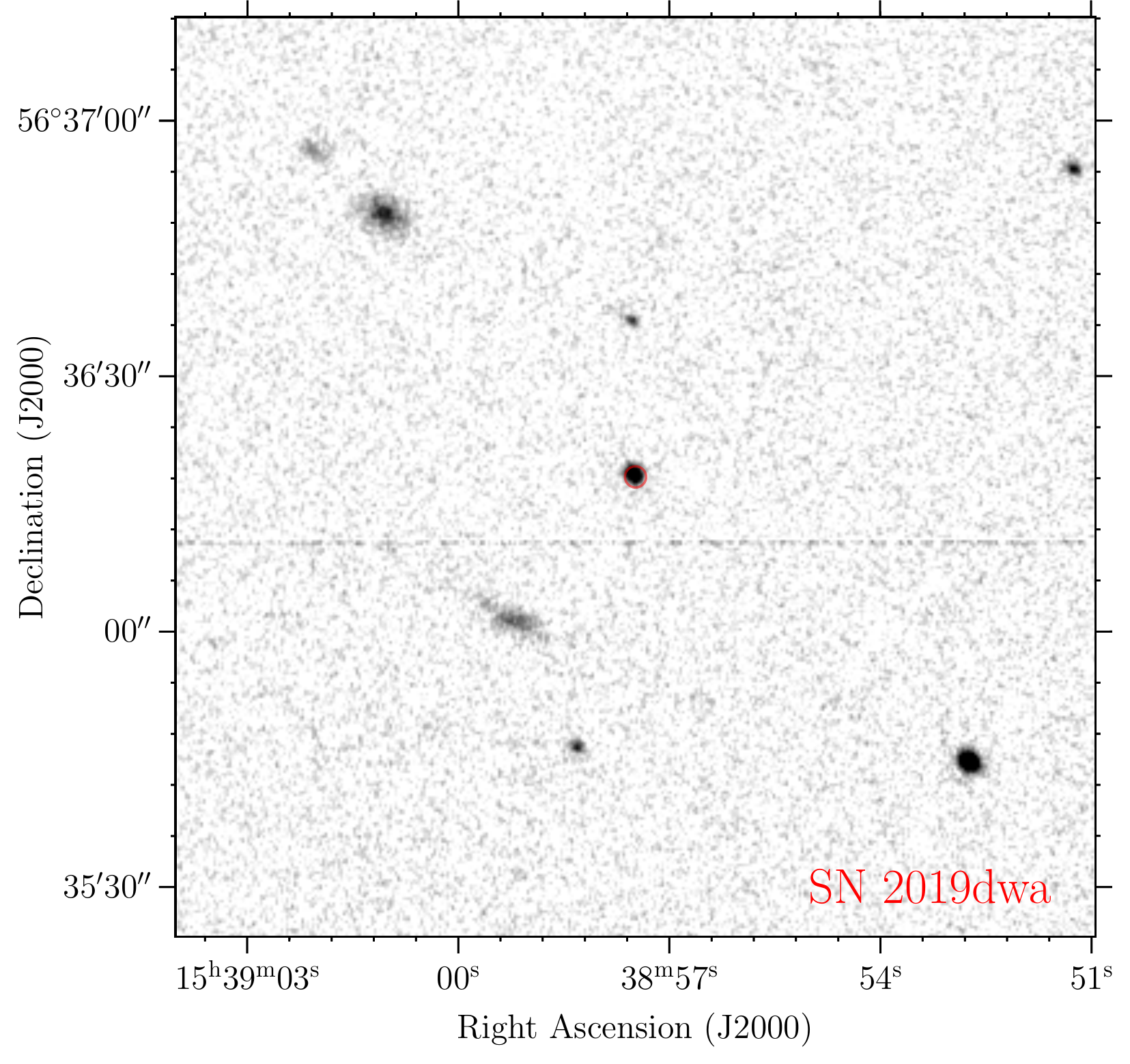}
    \includegraphics[scale=0.45]{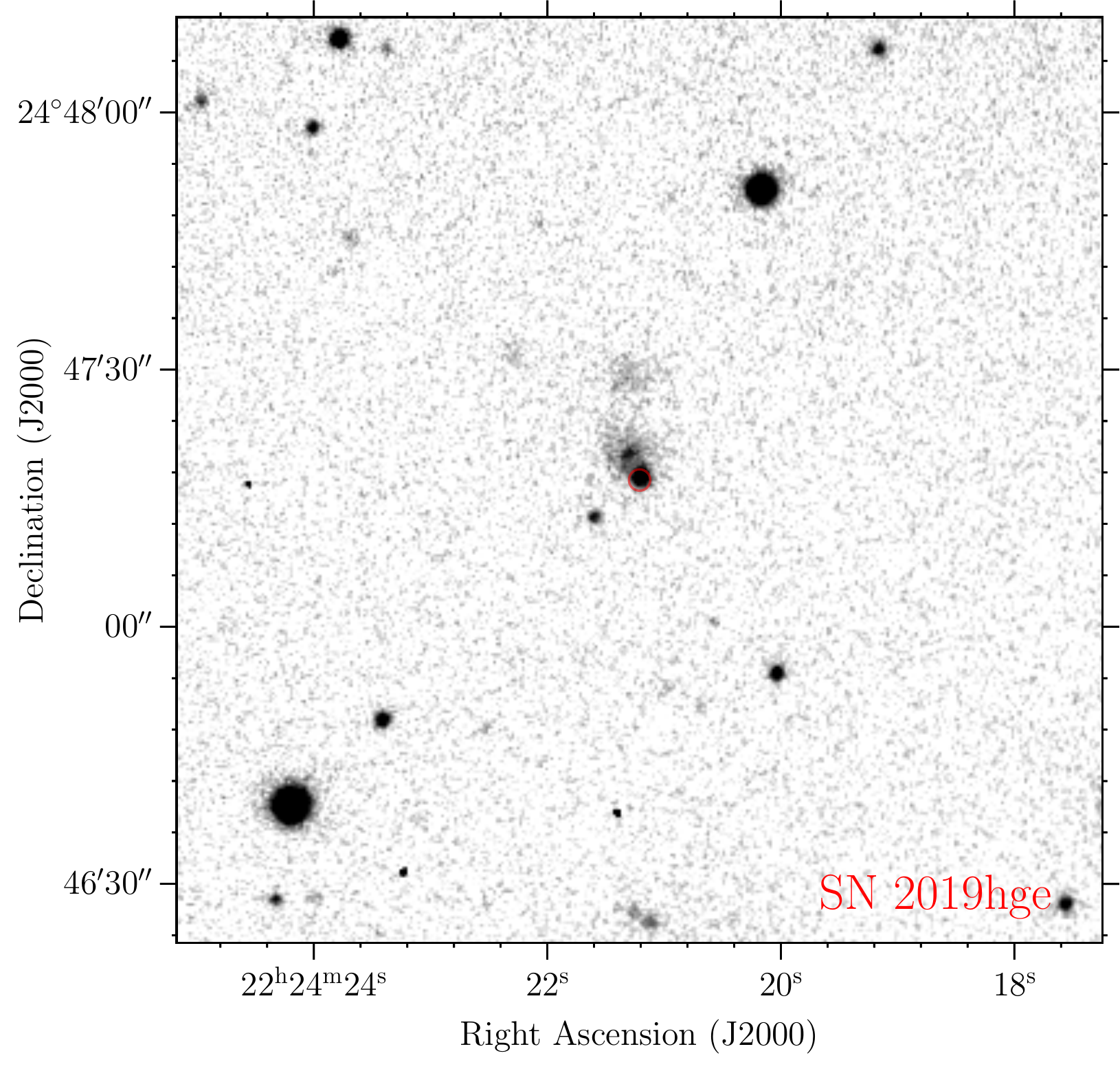}
    \includegraphics[scale=0.45]{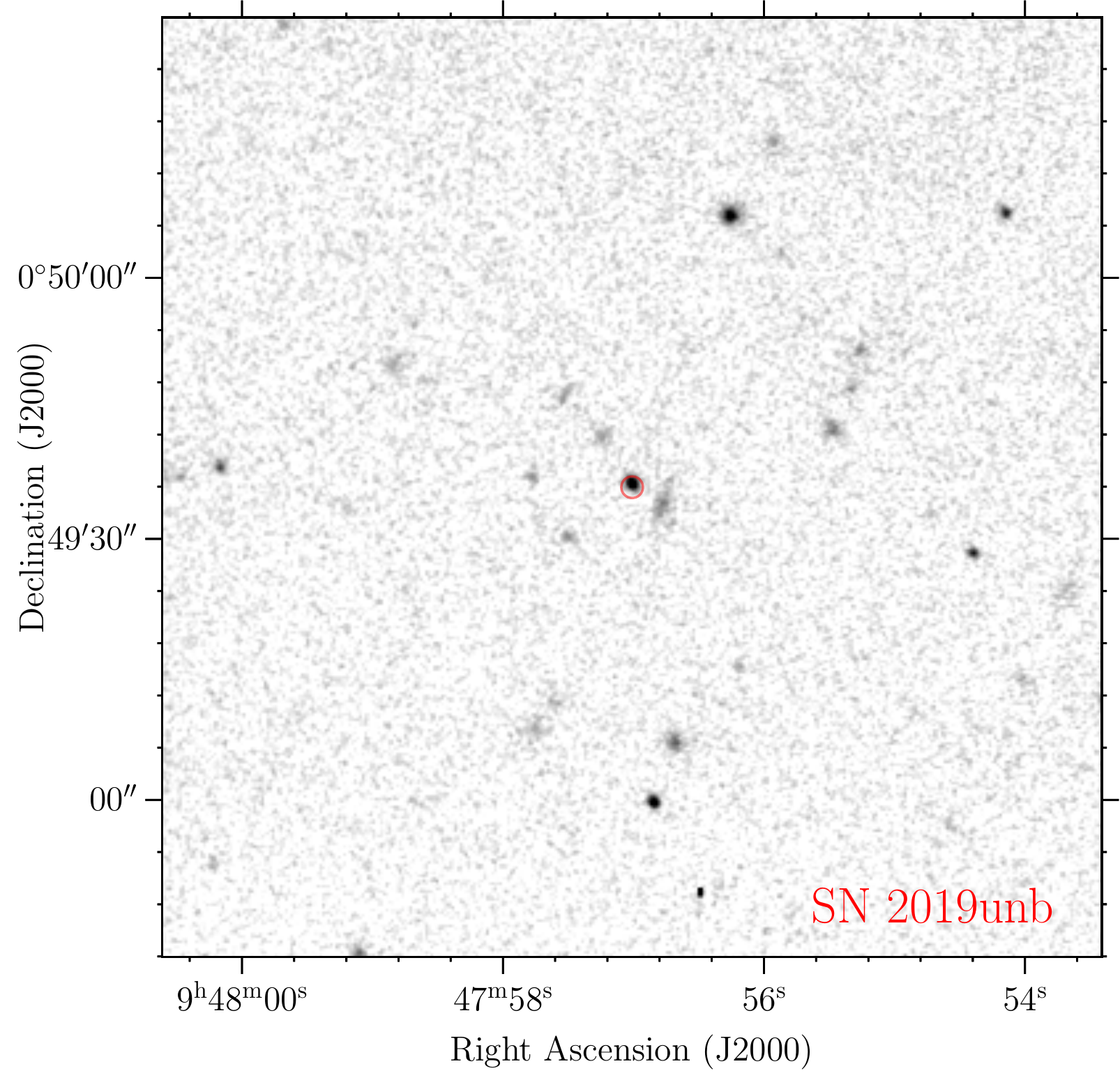}
    \caption{Liverpool Telescope $r$-band images of the \sn2019cri (upper left), \sn2019dwa (upper right), \sn2019hge (lower left), and \sn2019unb (lower right)   as observed in their respective fields. Owing to the distances involved, the host galaxies are small and dim, and at the location of each transient.}
    \label{fig:context}
\end{figure*}

The SNe, shown in context in Fig.~\ref{fig:context}, were observed with the Liverpool Telescope \citep[LT;][]{Steele2004}. Optical imaging was obtained with the IO:O camera and spectra with the SPectrograph for the Rapid Acquisition of Transients \citep[SPRAT;][]{Piascik2014}.
Optical photometry was reduced with a custom {\sc python} script utilising {\sc pyraf}, which was then calibrated to Sloan Digital Sky Survey \citep[SDSS;][]{Ahn2014} stars in the respective field.
The SPRAT spectra were reduced and calibrated using the LT:SPRAT pipeline \citep{2012AN....333..101B} and a custom {\sc python} script.

Three of the objects were observed as part of the advanced Public ESO Spectroscopic Survey of Transients \citep[ePESSTO+, for an overview of PESSTO see][]{2015A&A...579A..40S}. These are SNe 2019cri, 2019hge, and 2019unb.
For the aforementioned objects, spectra were obtained using ESO Faint Object Spectrograph and Camera (v.2) \citep[EFOSC2;][]{Buzzoni1984} on board the ESO New Technology Telescope. These spectra were reduced through standard pipelines\footnote{https://github.com/svalenti/pessto}. 

We were able to obtain additional spectra using the Intermediate-dispersion Spectrograph and Imaging System (ISIS) mounted on the 4.2\,m William Herschel Telescope (WHT). This was reduced using a custom {\sc IRAF} pipeline.

Finally, all proprietary data will be made public via the Weizmann Interactive Supernova Data Repository (WISeREP)\footnote{www.wiserep.org}

\subsubsection{Distance and extinction}
The distance modulus $\mu$ for each SN was calculated using its redshift, and cosmological parameters from the Nine-year Wilkinson Microwave Anisotropy Probe; $H_0=69.32$ km s$^{-1}$ Mpc $^{-1}$, $\Omega_m=0.286$, $\Omega_\Lambda=0.714$ \citep{Hinshaw2013}.
Reddening within the Milky Way, \Emw, were provided by the dust maps of \citet{Schlafly2011}.
The spectra of each object were checked for evidence of absorption by local dust via \NaI\ D absorption lines but no indication was found, thus we take intrinsic reddening \Eh\ for all objects to be negligible.
Relevant values are given in Table~\ref{tab:snproperties}.
All data were corrected for \Emw\ using $R_V=3.1$ and the extinction law of \citet{CCM}.

\begin{table*}
    \centering
    \caption{Redshift, extinction and distance of each object. Host extinction is assumed to be negligible.}
    \begin{tabular}{lccccccc}
    \hline
    SN & Type& $\alpha$ (J2000) & $\delta$ (J2000)  & $z$ &  $\mu$  & \Emw\ & Comments  \\
        &  &    &  &   &   [mag]   & [mag] &  \\
    \hline
    2019cri & Ic-7 (peculiar)    & 13:54:02.720  & +14:43:46.96   & 0.041  & 36.31  & 0.02 & Redshift from host \Ha    \\
    2019dwa & Ic-7 (peculiar)    & 15:38:57.480  & +56:36:18.18  & 0.076  & 37.71  & 0.01  & Redshift from host \Ha   \\ 
    2019hge & Ib-pec/SLSN-Ib    & 22:24:21.210  & +24:47:17.12  & 0.087  & 38.01  & 0.06 & Redshift from \citet{Yan2020}     \\
    2019unb & IIb-pec/SLSN-IIb    & 09:47:57.010  & +00:49:35.94  & 0.064  & 37.31  & 0.11 & Redshift from \citet{Yan2020}   \\
    \hline
    \end{tabular}
    \label{tab:snproperties}
\end{table*}

\subsection{Host Galaxies}\label{sec:hosts}

\begin{figure*}
    \centering
    \includegraphics[scale=.35]{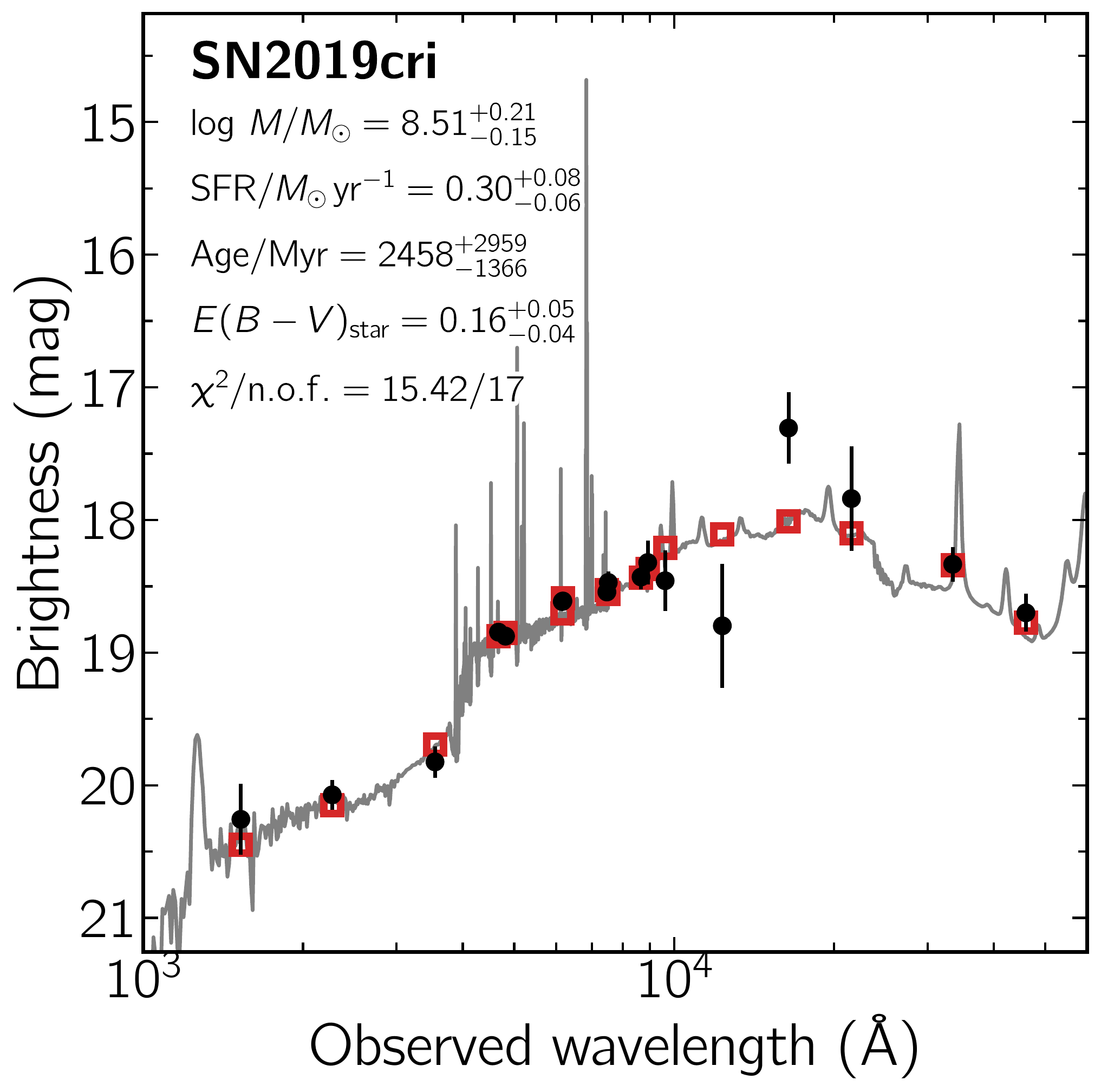}
    \includegraphics[scale=.35]{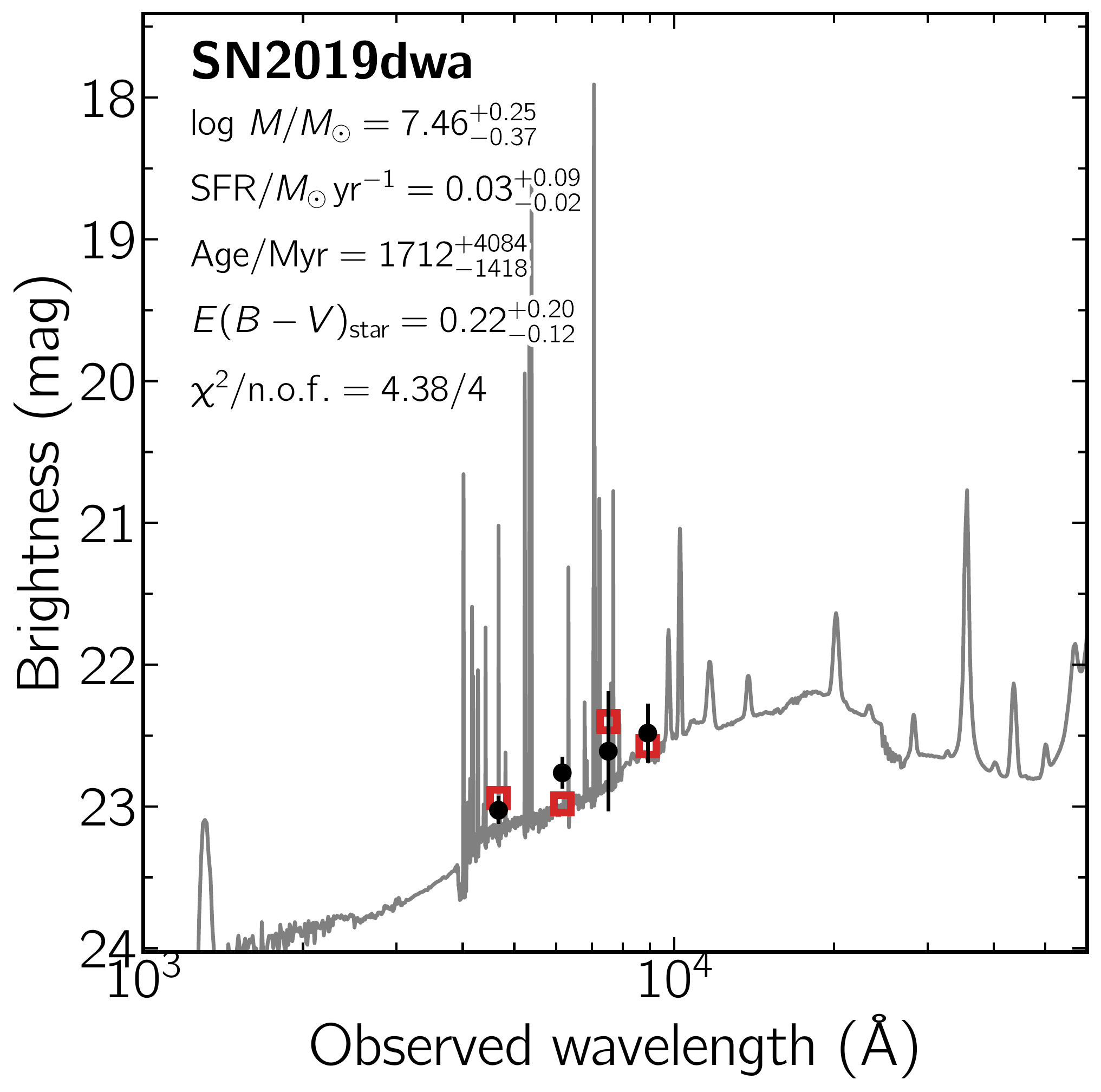}
    \includegraphics[scale=.35]{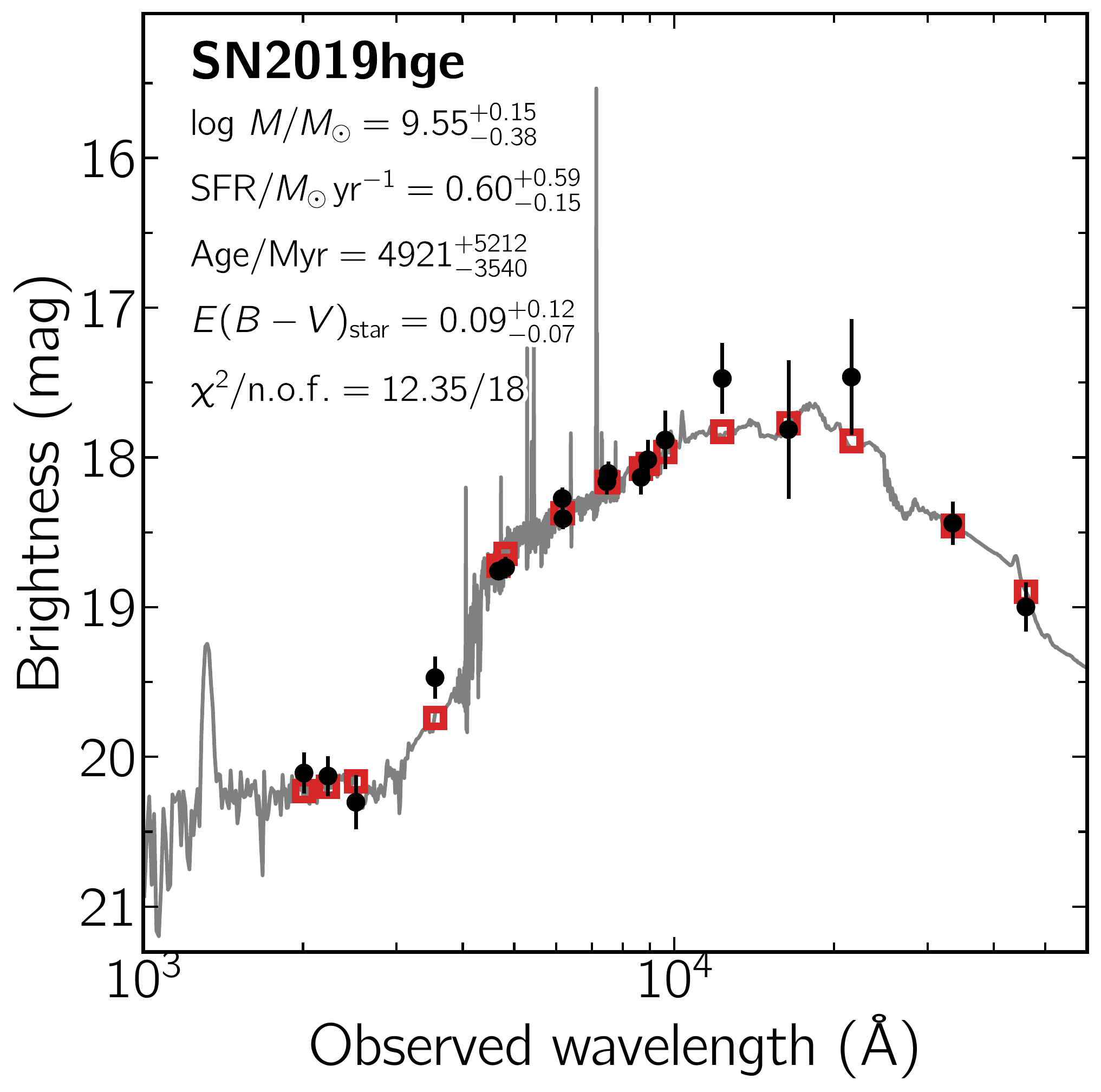}
    \includegraphics[scale=.35]{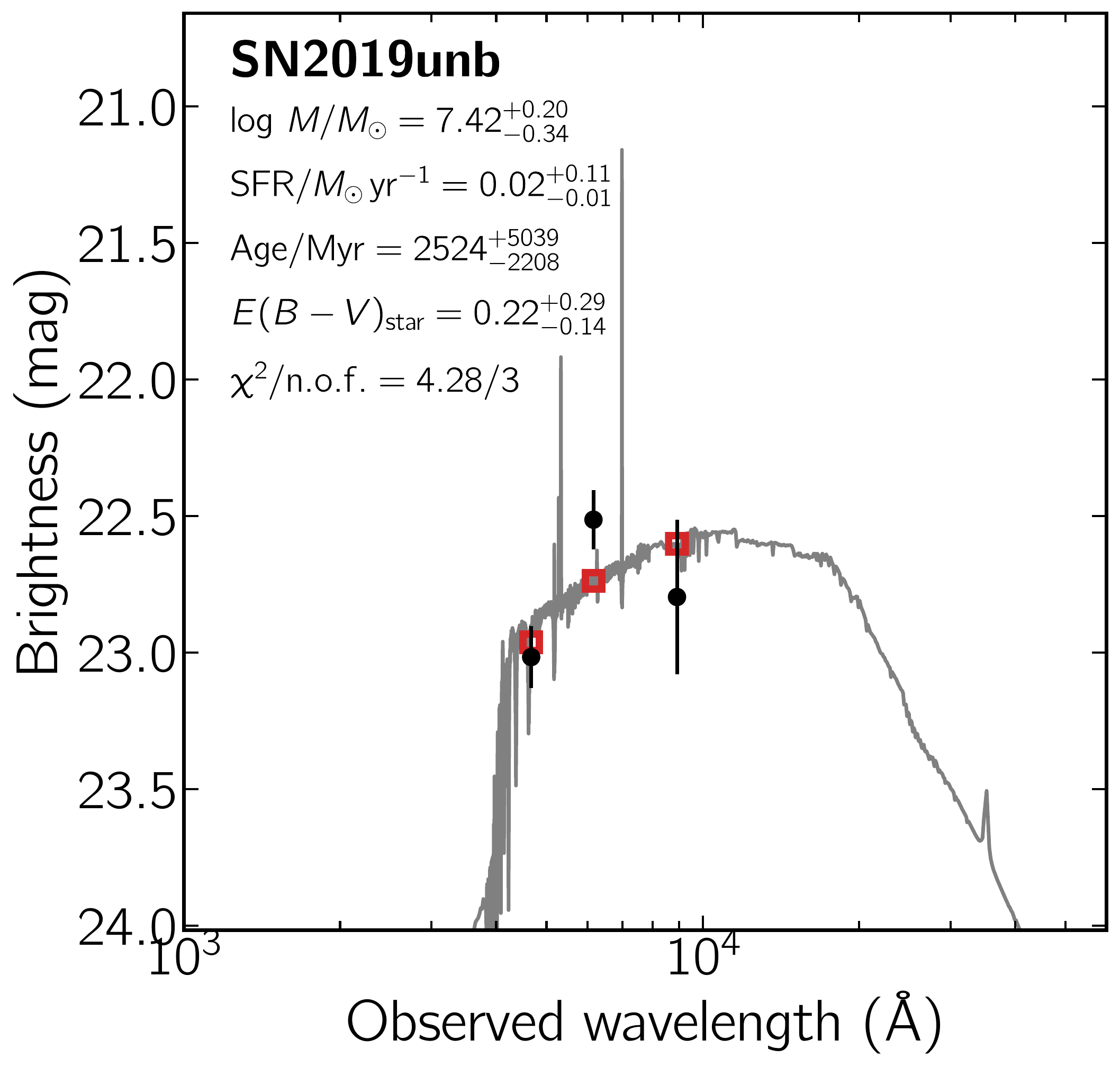}
    \caption{The spectral energy distribution of the host galaxies (detections $\bullet$; model predicted magnitudes $\square$). The solid line displays the best-fitting model of the SED. The fitting parameters are shown in the upper-left corner. The abbreviation ``n.o.f.'' stands for numbers of filters.}
    \label{fig:hosts}
\end{figure*}

\begin{figure*}
    \centering
    \includegraphics[scale=.47]{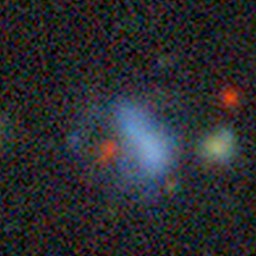}
    \includegraphics[scale=.47]{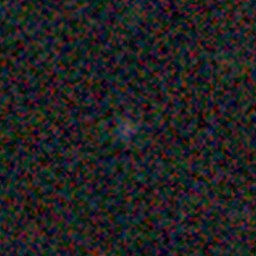}
    \includegraphics[scale=.47]{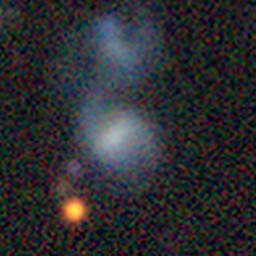}
    \includegraphics[scale=.47]{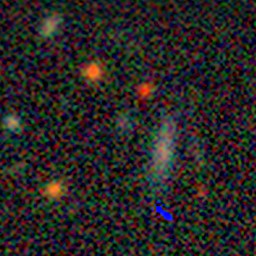}
    \caption{$grz$ cut-outs from The Dark Energy Camera Legacy Survey (DECaLS). {\bf Left:} The host of \sn2019cri. {\bf Centre left:} The host of \sn2019dwa. {\bf Centre right:} The host of \sn2019hge. {\bf Right:} The host of \sn2019unb. The morphological shape of the hosts of SNe 2019cri and 2019hge provide evidence of recent interaction. }
    \label{fig:decal}
\end{figure*}

The spectral energy distributions (SEDs) of the host galaxies were modelled with the software package {\sc Prospector}\footnote{\href{https://github.com/bd-j/prospector}{https://github.com/bd-j/prospector}} version 0.3 \citep{Leja2017a}. We assumed a linear-exponential star-formation history, the \citet{Chabrier2003a} IMF, the \citet{Calzetti2000a} attenuation model, and the \citet{Byler2017a} model for the ionized gas contribution. The priors were set as described in \citet{Schulze2020a}.
The results are shown in Fig.~\ref{fig:hosts}.

In order to build the SEDs, we retrieved science-ready coadded images from the \textit{Galaxy Evolution Explorer} ($GALEX$) general release 6/7 \citep{Martin2005a}, the Sloan Digital Sky Survey DR 9 (SDSS; \citealt{Ahn2012a}), the Panoramic Survey Telescope and Rapid Response System (Pan-STARRS, PS1) DR1 \citep{Chambers2016a}, the Two Micron All Sky Survey \citep[2MASS;][]{Skrutskie2006a}, and preprocessed $WISE$ images \citep{Wright2010a} from the unWISE archive \citep{Lang2014a}\footnote{\href{http://unwise.me}{http://unwise.me}}. The unWISE images are based on the public $WISE$ data and include images from the ongoing NEOWISE-Reactivation mission R3 \citep{Mainzer2014a, Meisner2017a}. We also retrieved deeper optical images from the DESI Legacy Imaging Surveys \citep[Legacy Surveys, LS;][]{Dey2018a} DR7 for \sn2019dwa and \sn2019unb. 
Co-added $grz$ images of the hosts from The Dark Energy Camera Legacy Survey (DECaLS) obtained through the Legacy Survey can be seen in Fig.~\ref{fig:decal}.

The field of \sn2019hge was observed with the UV/optical (UVOT; \citealt{Roming2005a}) on board the Neil Gehrels $Swift$ Observatory \citep{Gehrels2004a}. We use the data, after the SN faded, to measure the brightness of the host in the UV filters. The brightness in the UVOT filters was measured with UVOT-specific tools in the HEAsoft\footnote{\href{https://heasarc.gsfc.nasa.gov/docs/software/heasoft/}{https://heasarc.gsfc.nasa.gov/docs/software/heasoft/}} version 6.26.1. Source counts were extracted from the images using a region of $6.8''$. The background was estimated using a circular region with a radius of $39''$ close to the SN position. The count rates were obtained from the images using the $Swift$ tool {\sc uvotsource}. They were converted to magnitudes using the UVOT calibration file from September 2020. All magnitudes were transformed into the AB system using \citet{Breeveld2011a}.

We used the software package LAMBDAR (Lambda Adaptive Multi-Band Deblending Algorithm in R) \citep{Wright2016a}, which is based on a software package written by \citet{Bourne2012a} and tools presented in \citet{Schulze2020a}, to measure the brightness of the host galaxy. The brightness in the Legacy Surveys images were measured with the aperture-photometry tool presented by \citet{Schulze2018a}\footnote{\href{https://github.com/steveschulze/Photometry}{https://github.com/steveschulze/Photometry}}.

\begin{figure}
    \centering
    \includegraphics[scale=0.55]{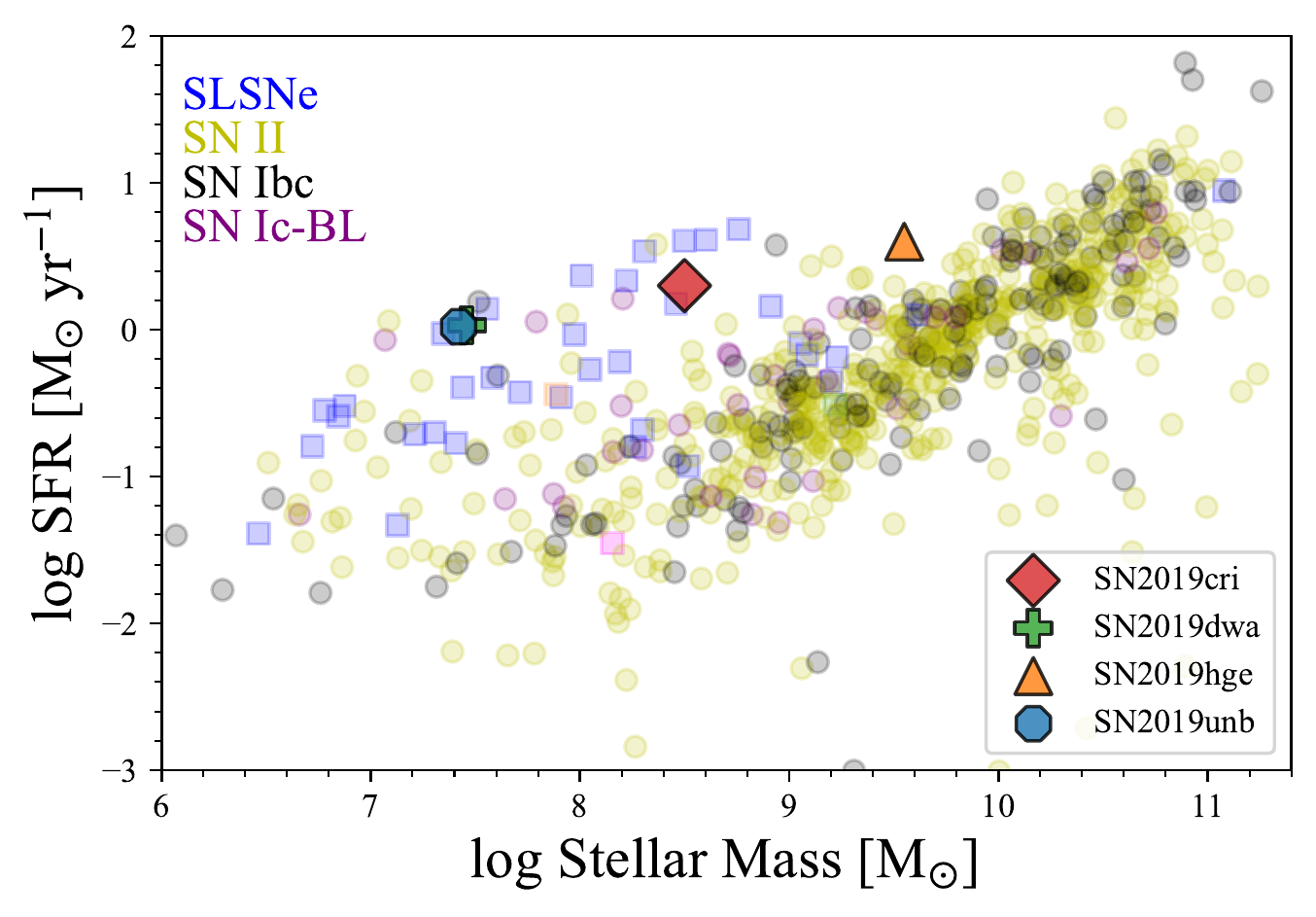}
    \caption{The position of the host of the four SNe in comparison with the PTF sample of \citet{Schulze2020}. While most of the host galaxies have high star formation and relatively low mass, the host of \sn2019hge stands out as being the most massive of the four, and comparable to regular CC-SNe. }
    \label{fig:hostprops}
\end{figure}

Figure~\ref{fig:hostprops} shows how these host galaxies compare with the hosts of the PTF sample of supernovae in star formation rate/stellar mass space. 
It can be seen that the hosts of SNe 2019cri, 2019dwa, and 2019unb have typically higher star formation rates than is usual for the hosts of `normal' CC-SNe, and are consistent with the hosts of SLSNe. In terms of stellar mass, the host of \sn2019hge has the largest and is comparable with the typical hosts of CC-SNe.
Finally, by inspecting the host galaxy morphology in the Legacy Survey images, we can see that the hosts of \sn2019cri and \sn2019hge have undergone, or are still undergoing, interaction with another galaxy, Fig.~\ref{fig:decal}.

\section{Photometry}
\subsection{Light curves}\label{sec:lcs}

\begin{figure*}
    \centering
    \includegraphics[scale=.65]{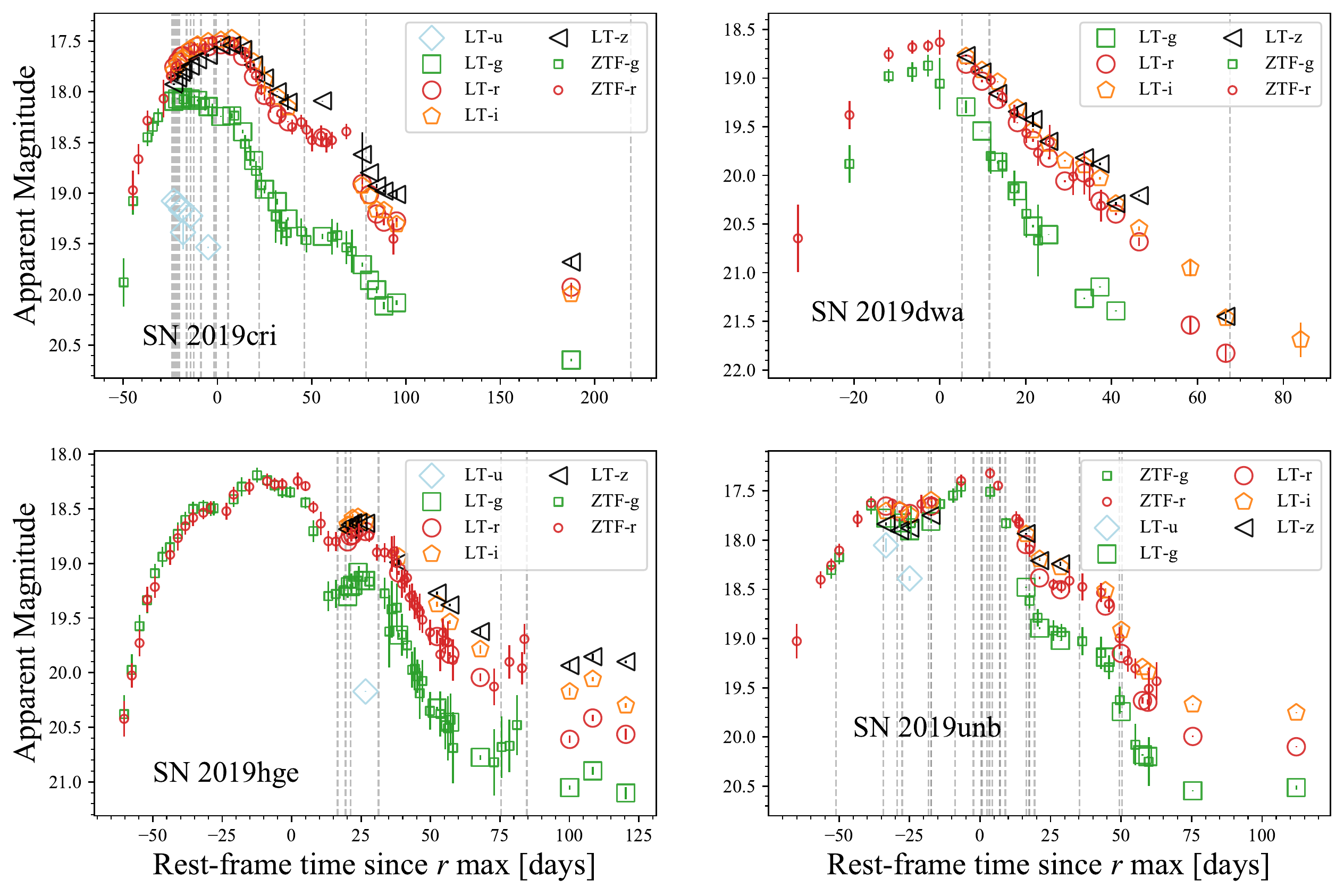}

    \caption{The multi-colour LT light curves of the four SNe featured here. Included in the figure is the public ZTF P60 photometry. {\bf Top left} \sn2019cri. {\bf Top right:} \sn2019dwa. {\bf Lower left:} \sn2019hge. {\bf Lower right:} \sn2019unb.   
    Grey dashed lines denote epochs of spectroscopic observations. }
    \label{fig:LCs}
\end{figure*}

Figure~\ref{fig:LCs} shows the multi-colour light curves for each object. Also plotted are the epochs of spectroscopic observations. For both \sn2019hge and \sn2019dwa, the unusual nature of the events were only publicly reported quite late in the evolution of each object, so our observations began post-maximum.
The presence of the public ZTF light curves are testament to the importance of public data release from large surveys, as without this data neither of these objects would have been identified as being unusual or been subject to a follow-up campaign.

From Fig.~\ref{fig:LCs} there are two obvious defining features for each event.
The first is the long lived rise to $r$-band maximum light. This is at least 65 days for \sn2019cri, 35 days for \sn2019dwa, 60 days for \sn2019hge and 70 days for \sn2019unb.
The second, are the variations in brightness prominently seen in \sn2019cri, \sn2019hge, and \sn2019unb. These variations are highly reminiscent of another recent event, SN2019stc, which was also identified as a transitional event between SLSNe and normal SE-SNe \citep{Gomez2021}.

\subsubsection{Light curve morphology}
The brightness variations are unlike the typical evolution of SE-SNe, although they are seen in SLSNe. If we consider \sn2019cri, the light curves rise to an initial peak after $\sim$ 60 days, before they decay but then level off between 80 -- 100 days. The light curves then undergo another period of decay, but this last little beyond 150 days because the final observations at $\sim 240$ days show little change from the observations almost 100 days prior.

\sn2019dwa shows the least variation in brightness, however, around 65 -- 70 days some variation is seen in $r$ that is present in both the LT and the ZTF photometry. A curious aspect of \sn2019dwa's light curve, aside from the long rise, is that after peak the light curve decays for nearly 100 days without ever settling on a \Cofs\ tail. This will be discussed further in relation to other objects in Section~\ref{sec:abs_r}. 

Next, we consider the complex light curves of \sn2019hge. The transient rises to a peak over 60 -- 70 days. Around 30 days the light curve appears to ``stall'' in its rise before continuing up to maximum light. At around 70 days the light curve drops 0.7 mag and 1 mag over about a week for $r$ and $g$ respectively.
The ZTF light curve then begins a slow rise again, at which point our photometric and spectroscopic observations start.
The rise lasts 12 days before decaying further, at a slightly slower rate than previously. At about 125 days the decay ceases and another rise is seen in the ZTF $g$ and $r$ photometry. This decays again before settling on a relatively flat tail. At this point the object was no longer visible during the night, it was not later recovered.

Finally \sn2019unb evolves in a similar way to \sn2019hge. There is a rise to a plateau over 30 days, the plateau is seen in the optical bands but not in $u$, which decays from the first observation. After $\sim55$ days the $griz$ light curves rise again until they reach maximum light at 70 days. The light curves then decay until around 95 days, with $g$ decaying more rapidly than $r$ as per \sn2019hge. There is then a period with much slower decay before the light curves return to a rapid decay between 120 -- 150 days. At this point the light curve is sparsely sampled but shows a very flat evolution. The photometry is not contaminated by the host and was the transient was clearly detected in the images on these days.

\subsubsection{Absolute $r$-band light curves}\label{sec:abs_r}
To estimate the absolute $r$-band light curves of the SNe, the relevant light curve was corrected for \Emw\ as per Table~\ref{tab:snproperties} then a K-correction was applied. These were derived from the spectra and assumed to be constant (the same as the calculated value nearest in time) outside of the epoch of spectroscopic observations. While this does miss the time-dependence of the K-correction at these phases the overall corrections are relatively small ($\sim |0.1|$ mag) and applied at phases away from maximum light.
Most affected is \sn2019dwa, where the $g$-band K-corrections approach $0.3$ mag and the spectroscopic time-series only covers a few weeks.
Finally, the absolute magnitude was found by subtracting the distance modulus $\mu$ from the corrected apparent magnitude. The peak $M_r$ and $M_g$ are given in Table~\ref{tab:peaks}.

\begin{figure}
    \centering
    \includegraphics[scale=.55]{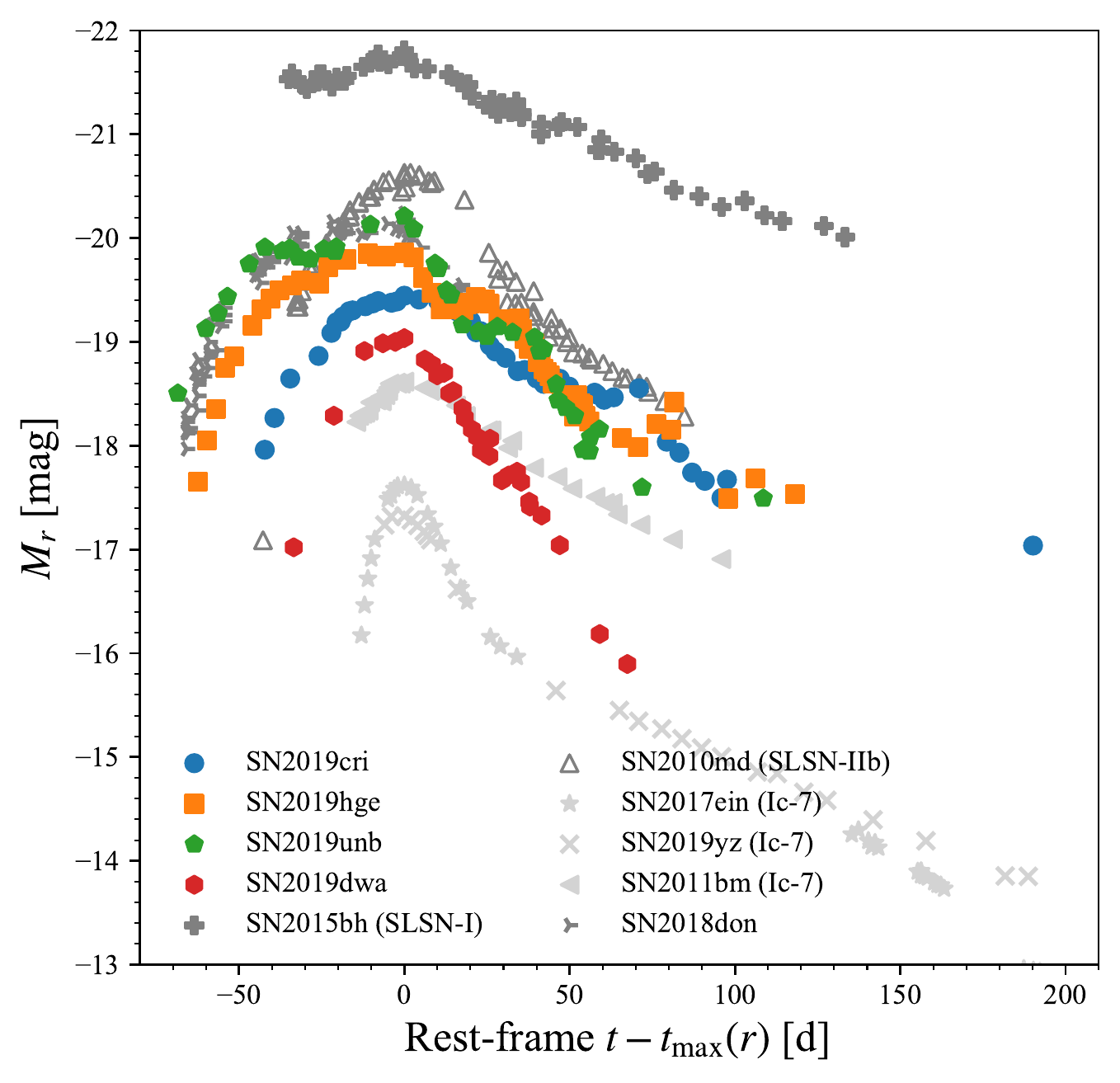}
    \caption{The absolute $r$-band magnitude of the four SNe. For comparison are a sample of SNe Ic-7 - SNe 2011bm \citep{Valenti2011}, 2017ein \citep{Teffs2021}, and 2019yz. SNe Ibc are generally found in the region bounded by these objects. Also included is a representative SLSN-I \citep[SN 2015bn][]{Nicholl2016}, as well as SLSN-IIb 2010md \citep{Quimby2018} and luminous SN 2018don. The four SNe presented here lay between the two supernova groups, in a region known for the lack of H-poor CC-SNe. Despite the classification of SNe 2019hge and 2019unb as SLSNe, only SN 2019unb has $M_r < -20$ mag, and this only fleetingly. }
    \label{fig:abs_r}
\end{figure}

Figure~\ref{fig:abs_r} shows these light curves against a selection of SE-SNe and SLSNe.
The four transients have peak $M_r$ between $-19$ and $-20.1$ mag, which places them in between the space of the luminous SNe Ibc and the sub-luminous SLSNe. 
\sn2019dwa is the least luminous of the four in $M_r$ and also has the narrowest light curve, as defined by the length of time the light curve is more luminous than half its peak luminosity (full width half maximum; FWHM).

Brighter still is \sn2019cri, \sn2019hge, and then \sn2019unb. 
As is discussed in Section~\ref{sec:spectra}, SNe 2019cri and 2019dwa are spectroscopically similar to normal SNe Ic, while SNe 2019hge and 2019unb are similar to SNe Ib/IIb at later phases but at early phases have a SLSN-like blue spectrum.
The well studied SLSN \sn2015bn \citep{Nicholl2016} peaks at nearly two orders of magnitude greater than our four objects, which are closer to the luminosity distribution of SE-SNe.

It was previously noted that \sn2019dwa has an unusual decline that does not reach a \Cofs\ tail by at least 70 days after maximum. Comparison with the SNe Ibc shows how unusual this is, as the decaying light curve of \sn2019dwa passes through the position of the late linear tails of many objects that are dimmer at peak magnitude.

Prior to the discovery of \sn2019cri, The Type Ic-7 \sn2011bm \citep{Valenti2012} displayed the broadest known light curve for a spectroscopically normal SNe Ic. Figure~\ref{fig:abs_r} demonstrates that \sn2019cri is a considerably broader and more luminous an event. The spectroscopic similarity between these SNe Ic-7 shown in the figure is discussed in Section~\ref{sec:compspec}.
\sn2019cri is also similar photometrically to the luminous H/He-poor SN 2018don \citep{Lunnan2020} and transitional object SN 2012aa \citep{Roy2016}, all display a long rise followed by a decline, a period of levelling off and then a secondary decline. Spectroscopically however, neither \sn2018don or \sn2012aa display the strong narrow absorption of \sn2019cri.

\begin{table}
    \centering
    \caption{Intrinsic peak magnitude and full width at half maximum light for $M_g$ and $M_r$}
    \begin{tabular}{ccccc}
    \hline
    SN & $M_g$ & Width  & $M_r$ & Width  \\
       & [mag]  & [days] & [mag]  & [days] \\
    \hline      
    2019cri &  $-19.02\pm{0.02}$  & 64   & $-19.45\pm{0.03}$  & 69     \\
    2019dwa &  $-19.0\pm{0.1}$  & 29   & $-19.0\pm{0.1}$   & 39    \\
    2019hge & $-19.98\pm{0.06}$   & 57   & $-19.86\pm{0.07}$   &  83    \\
    2019unb &  $-20.2\pm{0.1}$  & 64   & $-20.21\pm{0.05}$  &  67   \\
    \hline
    \end{tabular}
    
    \label{tab:peaks}
\end{table}

\subsection{Colour curves}
\begin{figure}
    \centering
    \includegraphics[scale=.55]{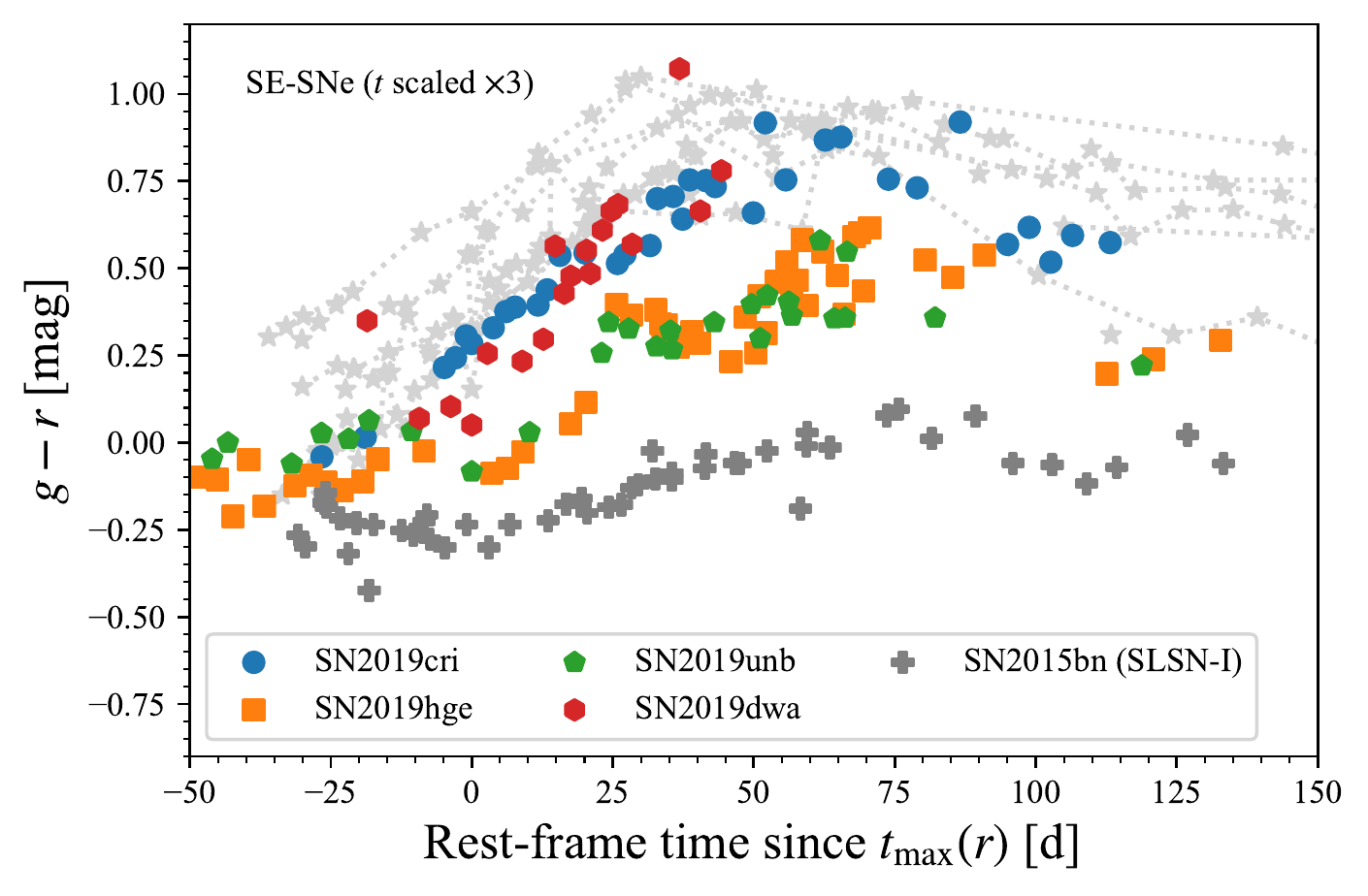}
    \caption{$g-r$ colour curves for the four transients, and for comparison the colour curves of a sample of SE-SNe (light grey) and SLSN 2015bn (dark grey). The four SNe and SN 2015bn have had K-corrections applied.  }
    \label{fig:cc}
\end{figure}

The $g-r$ colour curves derived from the dereddened and K-corrected $g$ and $r$ photometry are shown in Fig.~\ref{fig:cc}.
For comparison are a sample of SNe Ibc from \citet{Prentice2019} and a sample of SLSNe. 
Owing to the difference in temporal evolution between the SNe Ibc and the comparison object, the timescales of the SNe Ibc are multiplied by a factor of three.

SNe 2019cri and 2019dwa, which are both He-poor, follow the scaled evolution of the SNe Ibc. 
They both initially evolve slowly to the red, with \sn2019cri reaching a turnover at the same time as the scaled SE-SN colour curves. This is not apparent in the colour curve of \sn2019dwa and reflects that its light curves are never observed to settle on a late linear tail. 
In SE-SNe, the epoch of this turnover defines the early nebular phase when emission lines start appearing in the spectra, and this is also the case for \sn2019cri. 
The two SNe are bluer than the average for the SE-SNe, a slightly lower scaling would better fit the colour curve of \sn2019dwa but it does serve to show that the scaled evolution of the objects are similar.
The relationship between colour curves and reddening is well established \citep[see][]{Drout2011,2018A&A...609A.135S} and would suggest that the assumption of negligible \Eh\ is valid.

SNe 2019hge and 2019unb have very similar colour curves, especially in terms of the shape. They both start with a constant colour for some 50 days before evolving to the red, they then both take a series of blue-red-blue turn at similar times. 
They are bluer than the SE-SNe, but redder than the SLSNe but their colour curve shape is not similar to either, so no scaling will match the other objects in the same way as for SNe 2019cri and 2019hge.

\section{Spectroscopy}\label{sec:spectra}

The previous section demonstrated that these SNe have peculiar light curve evolution compared with SE-SNe and SLSNe. In this section the spectroscopic observations of each object will be discussed in turn after we first establish each object's similarity to a reference object.

\subsection{Comparison with other objects}\label{sec:compspec}
\begin{figure}
    \centering
    \includegraphics[scale=0.6]{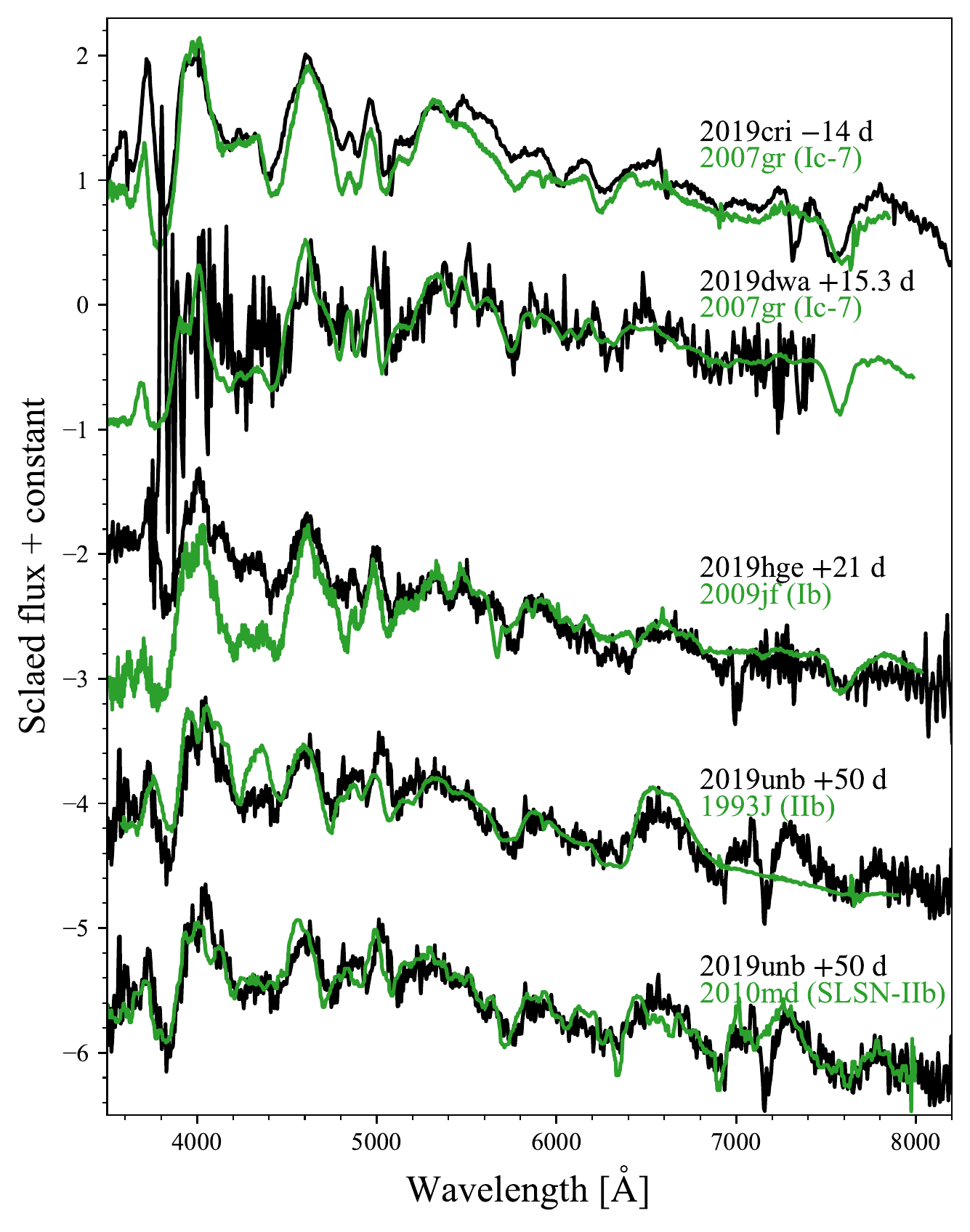}
    \caption{Spectroscopic comparisons between the four objects and spectra of SE-SNe in the photospheric phase. SNe 2019cri and 2019dwa match Ic-7 object, shown for comparison is \sn2007gr \citep{Hunter2009}. 
    SN 2019hge is matched against the Type Ib 2009jf, and \sn2019unb against Type IIb SN 1993J and SLSN-IIb 2010md. Additional matches can be found between these objects and most SNe Ib and H-weak SNe IIb. Note that the comparison spectra have been shifted slightly in velocity space to match the line velocities of the target objects.
    }
    \label{fig:compare_spectra}
\end{figure}

The classification and parameters of the SNe in question depend upon similarity with known transients and their properties. This applies in particular in the line identification.
Comparisons with various SE-SNe are shown in Fig.~\ref{fig:compare_spectra}.

Starting with the simplest case first, that of \sn2019cri. It can be seen in Fig.~\ref{fig:compare_spectra} that this transient has very similar optical spectra to Type Ic-7, which are the narrow line H/He-poor SE-SNe \citep{Prentice2017}. SNe in this sub-class include \sn2007gr \citep{Hunter2009}, which is shown for comparison, as well as \sn2011bm \citep{Valenti2012}, \sn2017ein \citep{Teffs2021}, \sn2005az, \sn2014L, and \sn2019yz\footnote{Data for this object is presented in conjunction with this work.}.
These SNe are characterised by having the lowest photospheric velocities of SNe Ic but the largest range of light curve morphologies and luminosities, which suggests that objects in this spectroscopic sub-type can have a large range of ejecta masses \mej\ and nickel masses \mni.

Next, we find that \sn2019dwa matches the spectra of several SNe Ic and SNe Ib, with a preference for the former. Our spectra are all more than a week after $r$-band maximum light, and SE-SNe at this phase become increasingly difficult to classify as their spectra tend to appear similar. The absence of strong \HeI\ lines at 6678 \AA\ and 7065 \AA\ would support that this is a SNe Ic. There is an absorption at 5800 \AA, but this is likely to be due to \NaI\ D rather than \HeI\ \lam5876. The comparison with Ic-7 \sn2007gr shows that this line is present in a similar strength in SNe Ic.

Finding comparison objects for \sn2019hge and \sn2019unb is considerably more difficult, as these are genuinely unusual objects spectroscopically. We attempted to match the early spectra, but no match with publicly available data was adequate.
We did find matches to SE-SNe at the later phases of each object; SN Ib 2009jf \citep{Valenti2011} for \sn2019hge, and Type IIb \sn1993J \citep{2000AJ....120.1487M} for \sn2019unb, these comparisons are shown in Fig.~\ref{fig:compare_spectra}. 
As can be seen, the features of the transients overlap, although \sn2009jf is redder than the others.
The \sn2009jf  and \sn1993J spectra are not plotted in the rest frame, a redshift was artificially introduced to match the line velocities seen in the other two objects. Fundamentally, this is unimportant however, as line velocities vary from object to object and this shift can be replicated by comparing spectra with small shifts in velocity space \citep[See][]{Mazzali2017}.
\sn2019unb also matched the SLSN-IIb 2010md \citep{Quimby2018,Shivvers2019} during its later phases, but it may also be said that \sn2010md matched He-rich SE-SNe in its later phases, and its peak luminosity was similar to that of \sn2019unb.
The failure to find spectroscopically similar objects at earlier phases, and the later similarity to SN~Ib/IIb, led to an exploratory attempt to replicate the early spectra of SNe 2019hge and 2019unb by convolving the pre-maximum spectra of several SE-SNe and a thermal continuum.
This led to some success in replicating features in the earlier spectra that could be identified as \CaII\ H\&K, \FeII, \HeI\ of \NaI\ D, and \Ha\ or \SiII. This is investigated in more detail in Section~\ref{sec:19unb_model}.

\subsubsection{Classification}
As has been demonstrated, \sn2019cri is spectroscopically similar to SNe Ic-7 and we retain this classification. 
Likewise, from limited spectroscopy of \sn2019dwa we find more of a match to narrow line SNe Ic-7 rather than to SNe Ib.

There exists no neat box in which to place \sn2019hge and \sn2019unb, they could be classified as SN Ib-pec and SN IIb-pec respectively, where `pec' is short for `peculiar', or SLSN-Ib/IIb.  
It has been shown that these objects, while luminous, are not clearly at SLSN luminosities. Therefore they may represent a transitional form of event that in different circumstances could lead to either a normal SN IIb or a SLSN.

\subsection{The spectra of SN\,2019cri}\label{sec:19cri_spectra}
\begin{figure}
    \centering
    \includegraphics[scale=0.55]{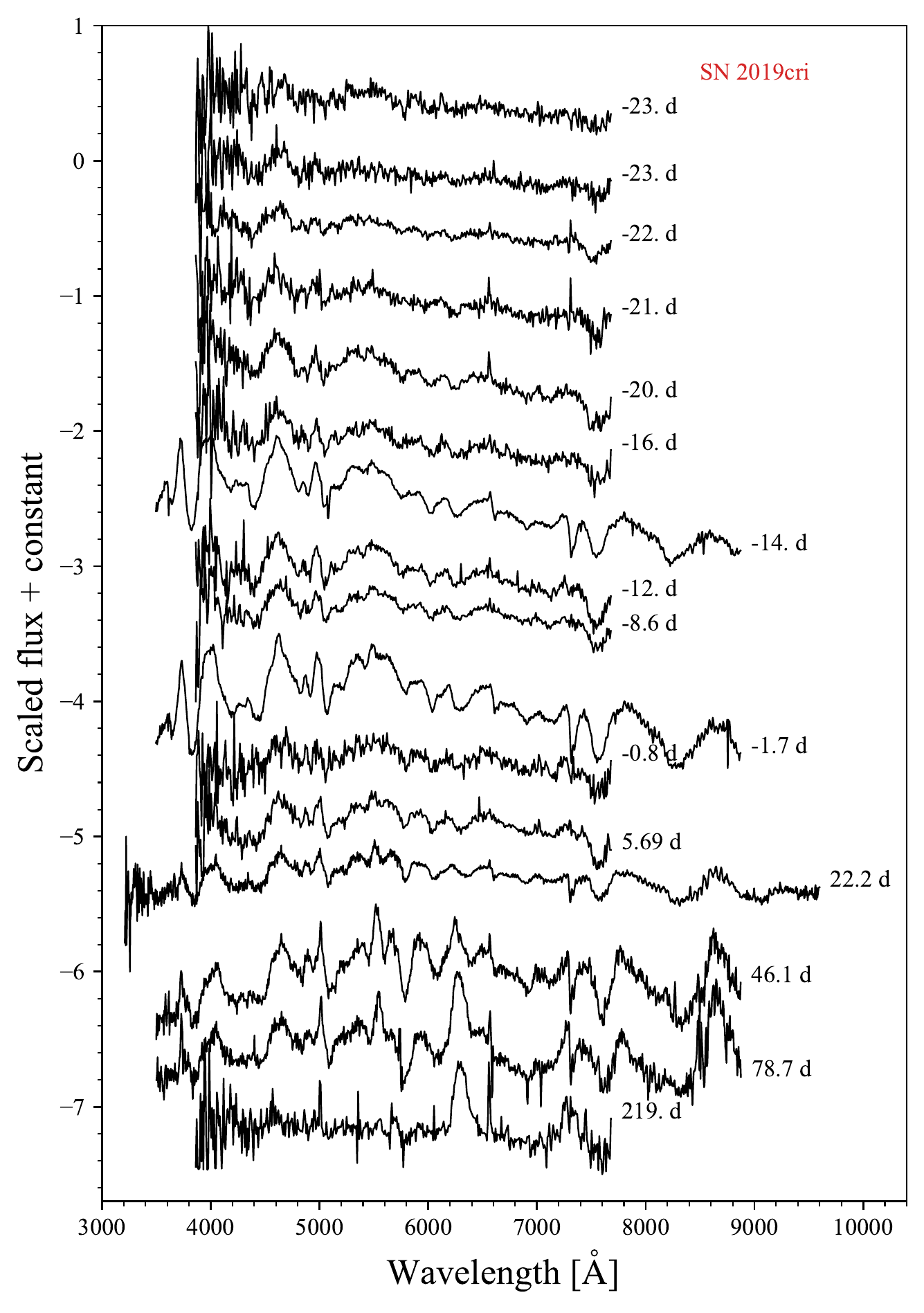}
    \caption{The spectroscopic sequence of SN 2019cri as observed by the LT and NTT. Despite the slow photospheric evolution, the SN enters the early nebular phase two months after maximum light, which is consistent with the timescales for normal SNe Ic. All phases are relative to the time of $r$-band maximum light.}
    \label{fig:19cri_spec}
\end{figure}

Figure~\ref{fig:19cri_spec} shows the spectroscopic observations of SN 2019cri. As previously discussed, the spectroscopic evolution is virtually identical to that of SNe Ic-7.
The most significant difference is an absorption feature around 6100 \AA. This is not usually seen in SE-SNe, and it may be \OI\ \lam6158.
The maximum light spectrum is modelled in Section~\ref{sec:19cri_model} in order to investigate this feature.

The \Oneb\ emission line is present in the spectra at $+46.1$ days, which would be a typical timescale for normal SE-SNe. This contrasts with the object's unusual photometric evolution, which would lead one to expect a later transition into the early nebular phase.
By +78.7 days, the \Oneb\ line is the dominant feature in the spectrum, also present is the \caiif\ emission line. 
The observation coincides with the time that the light curve begins to fall rapidly again.
At $+219$ days the spectrum is entirely nebular. Assuming that the entire emission line at $\sim6300$ \AA\ is from one \Oneb\  doublet, the FWHM of the [O~{\sc i}] \lam6300 component is $5000\pm{200}$ \kms. 
This is at the very low end of the FWHM velocity distribution for SNe (Prentice et al. submitted), and shows that the relatively low photospheric velocities are also seen in the nebular phase.
Finally, taking the flux ratio of \caiif/\Oneb\ gives $\sim 0.8$.

It is important to point out that there is no indication of H or He emission lines in the photospheric spectra, and there are no signatures within the spectra that mark the changes in the light curve. Spectroscopically, \sn2019cri is the same as any other Ic-7.

\subsection{The spectra of SN\,2019dwa}
\begin{figure}
    \centering
    \includegraphics[scale=0.55]{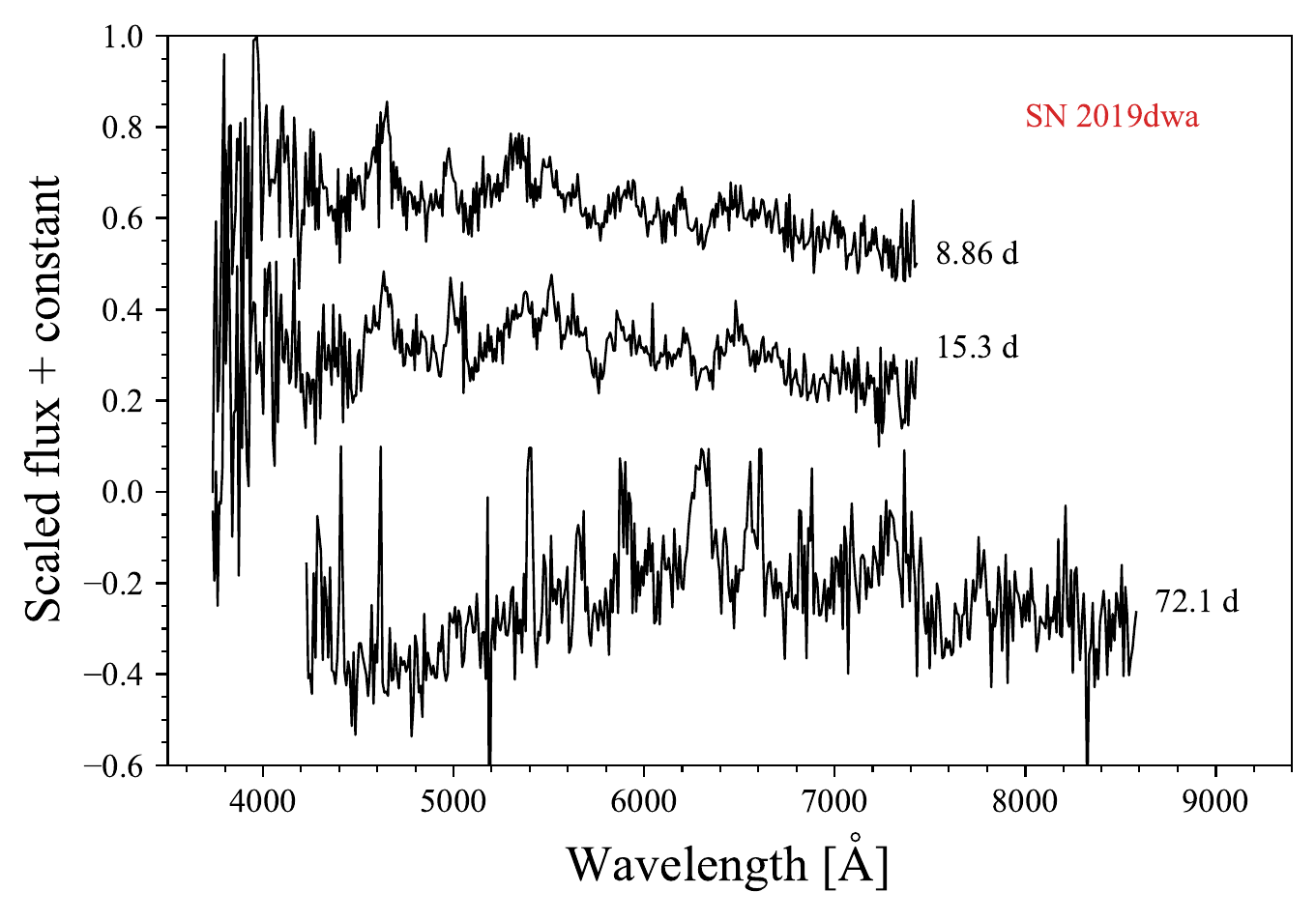}
    \caption{Spectroscopic observations of SN 2019dwa. The earliest two spectra are from the Liverpool Telescope. The final spectrum is from the WHT, and is affected by poor flux calibration. The important aspect of this spectrum is the presence of the \Oneb\ and \caiif\ lines. There is also a narrow \Ha\ line present from the host galaxy. These confirm the core-collapse origin of this object and set the redshift as $z= 0.076$.
    }
    \label{fig:19dwa_spec}
\end{figure}

\sn2019dwa was too far North to be observed by ePESSTO+, so we obtained just three spectra for this object; two shortly after one another about a week after $r$ maximum. Figure~\ref{fig:19dwa_spec} shows that they are of low signal to noise but provide enough information to say that the spectra are not H or He rich and there are no signs of emission lines.
The earliest spectra, taken just after $r$-band maximum, match well with that of SN~Ic-7. 

Two spectra were taken of the transient at $+72.1$ d using the WHT. The combined spectrum is noisy, and we reject any features that are not present in both spectra, but we can see that the dominant emission at this time is from \Oneb, also present is \caiif\ and \NaI\ D.
A full-width half-maximum measurement of the [{\sc O~i}] \lam6300 component to the doublet is $5000 \pm{300}$ \kms. 
Like \sn2019cri, this is consistent with measurements of SE-SNe \citep{Taubenberger2009} but is lower than the average.
We also detect \Ha\ in emission, and an emission line at 5670 \AA, which could be [N~{\sc ii}] \lam 5680.
The presence of these lines provides a redshift of $z = 0.076\pm{0.02}$ for the transient.
The \Ha\ line is noisy but thin, which suggests that it is emission from the host galaxy, this is discussed further in Section~\ref{sec:hosts} where this interpretation is at odds with the photometry of the pre-explosion site. If this is the case, then this line may be weak emission from interaction.

\subsection{The spectra of SN 2019hge}
\begin{figure}
    \centering
    \includegraphics[scale=0.55]{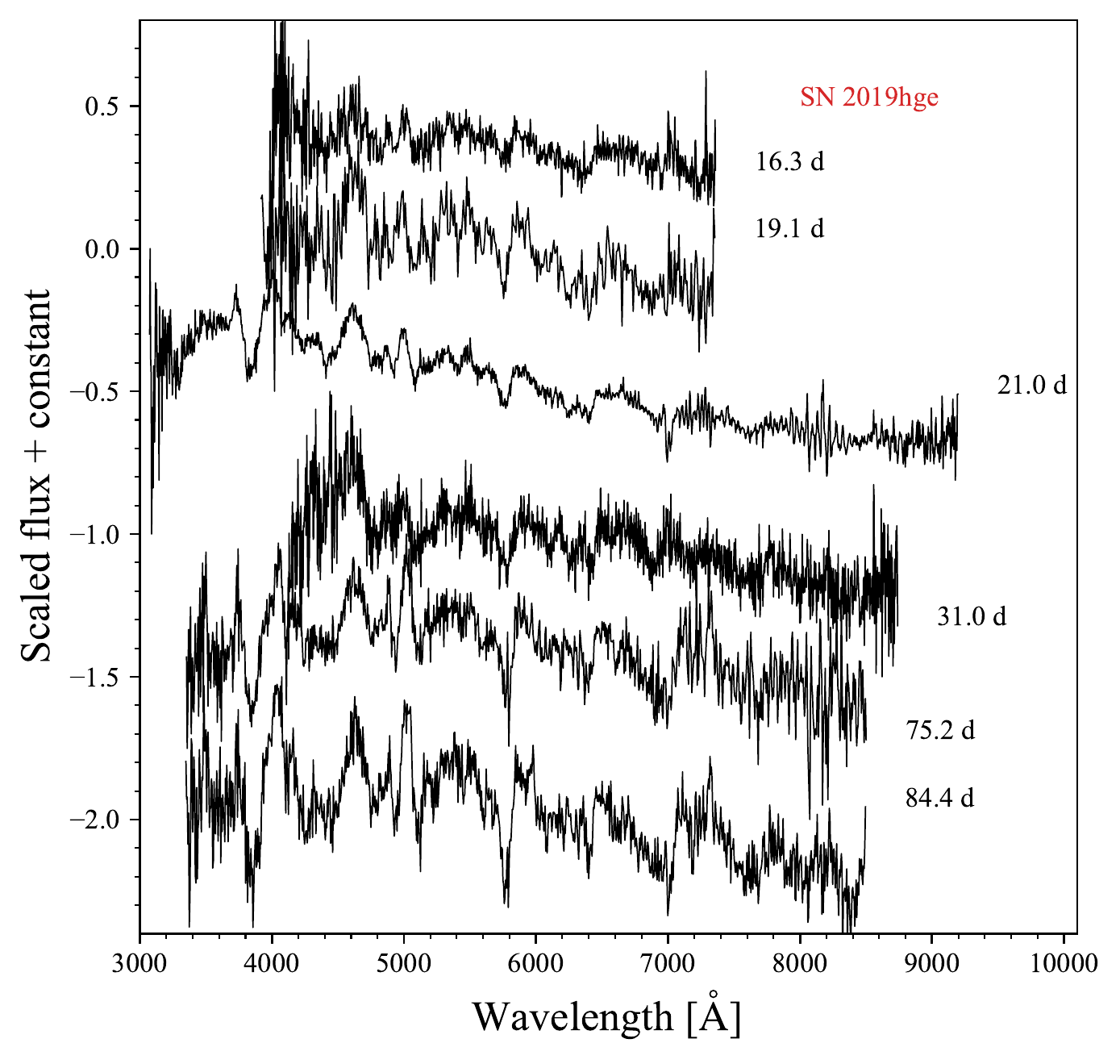}
    \caption{The spectra of \sn2019hge, phases are in the rest-frame and relative to $r$ maximum.}
    \label{fig:19hge_spec}
\end{figure}

Our observations of \sn2019hge only began after the secondary maximum, these a presented in Fig.~\ref{fig:19hge_spec}.
At these epochs the spectra are typically SN Ib-like, with features from \FeII\ \lam, \MgI\ \lam, \CaII\ H\&K, in the blue, \HeI\ \lam\lam5876, 6676, 7065, in the mid part of the spectrum, and \OI\ \lam 7774 and the \CaII\ NIR triplet in the red.
As with the previously discussed objects, there is no evidence for external influences on the spectra via IIn/Ibn like narrow emission lines that could explain the light curve variation.

The ZTF classification spectrum from MJD 58665.41 \citep{2019TNSCR2859....1D} is 30 days after discovery ($\sim48$ rest-frame days prior to our first spectrum) and coincides with the brief flattening in the light curves during the rise. 
This spectrum is blue, as to be expected from the colours. The colour curves also suggest that this spectrum changes little in terms of continuum slope for much of the rise. This spectrum does not display the characteristic `w' feature between 4000 -- 5000 \AA\ that is associated with \OII\ and is common in SLSN-I.

\subsection{The spectra of SN 2019unb}
\begin{figure}
    \centering
    \includegraphics[scale=0.55]{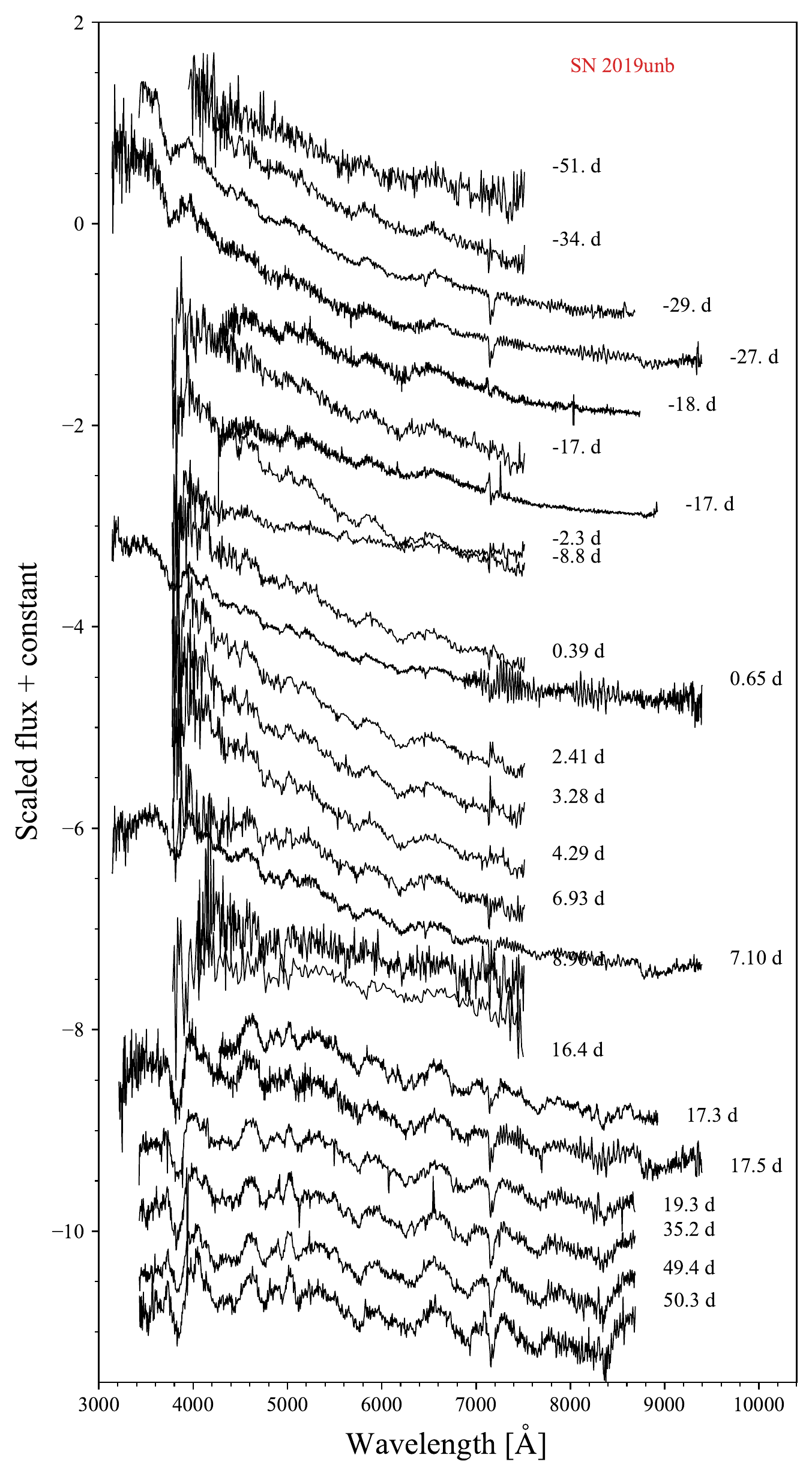}
    \caption{The spectroscopic sequence of \sn2019unb. Phases are relative to $r$ maximum.}
    \label{fig:19unb_spec}
\end{figure}

Unlike \sn2019hge, \sn2019unb was recognised early as an unusual event. Our first series of spectra (Fig.~\ref{fig:19unb_spec}) from $-50$ to $+7$ days are defined by a hot continuum with absorption present from \CaII\ H\&K, \HeI\ \lam5876 and/or \NaI\ D, and also possibly some weak Fe-group features between 4000--5000 \AA. 
As with \sn2019hge these are revealed through a simple convolution between the spectra of SE-SNe and a thermal continuum, but it is also acknowledged that the temperature would suggest different ionisation regimes for the various elements and it is possible that such features may be a chance alignment.
The spectra do not show the numerous \CII\ and \OII\ lines common to SLSNe, and as with the previous objects, we find nothing that evolves to trace the variations in the light curves and we find no obvious emission lines. The maximum light spectrum is modelled in Section~\ref{sec:19unb_model}.

Throughout its evolution, \sn2019unb displays a broad bump around 6563 \AA, which becomes more prominent over time and which may be \Ha. There is no measurable absorption component to this line however, but it does suggest the presence of H in the ejecta. This would also support the presence of He and thus the comparison with SNe Ib/IIb.

\subsection{Line velocities}\label{sec:line_vels}

To further characterise the properties of the SNe, photospheric phase line velocities were measured from the absorption minima of elements common to such transients. 
Figure~\ref{fig:velocities} shows the measurements for \FeII\ \lam5169, \SiII\ \lam6355, \CaII\ (due to limited spectroscopic wavelength, this is measured from both the NIR feature and H\&K), and \HeI\ \lam5876 where possible. Also presented for comparison are a small sample of SNe Ib/IIb \citep{Prentice2019} and a sample of SNe Ic-7, including \sn2011bm.

\subsubsection{\FeII\ \lam5169}
In terms of \FeII\ velocity, all four transients have significantly lower velocities that any of the comparison objects. \sn2019cri shows evolution over time, from $\sim 8000$ \kms\ to $\sim 5000$ \kms\, and evolves on a similar timescale to \sn2011bm.
If the velocity curve is scaled by 15/65, which represents the ratio of rise times for a normal SNe Ic compared to \sn2019cri, then the curve shapes are similar between it and the SNe Ic-7, but with \sn2019cri being $\sim2500$ \kms\ slower. That the velocity decreases until around optical maximum light and then levels off, is characteristic to all SNe Ic.
Only two measurements can be made of \sn2019dwa\ and both suggest that it also has low velocities at this phase.
\sn2019hge was observed well after peak, and displays an \FeII\ velocity commensurate with SNe Ib/IIb $\sim6000$ \kms. There is a reduction in this velocity over time to $\sim 3000$ \kms\ at $+80$ days.
\sn2019unb has the lowest maximum light \FeII\ velocity of all at $v \sim 4500$ \kms.
The \FeII\ lines are not strong however, so measurement is made by considering \FeII\ \lam\lam4924, 5018 as well. 
The measured velocities represent the only set of features that lined up with the strong series of \FeII\ lines between 4000 and 5000 \AA.
Comparison with \sn2009jf in Fig.~\ref{fig:compare_spectra} shows that the 4000--5000 \AA\ regions are similar at later times. In \sn2009jf, which is one of the comparison objects in Fig.~\ref{fig:velocities}, the \FeII\ velocities are easily traceable. 

\subsubsection{\SiII\ \lam6355}
This line is commonly measured in SNe Ic, it can be seen in Fig.~\ref{fig:velocities} that SNe 2019cri and 2019dwa have lower velocities than the SNe Ic-7. The maximum measured velocity for \sn2019cri is $\sim 6000$ \kms\ compared with 10\,000--12\,000 \kms\ for the comparison objects. Both \sn2019cri and \sn2019dwa show a post-maximum velocity of $\sim3000$ \kms, approximately $4000$ \kms\ slower than the SNe Ic.

\subsubsection{\CaII}
The velocity measurements of \CaII\ NIR and \CaII\ H\&K show the most similarity to the comparison SNe. SNe 2019hge and 2019unb both follow the same track as the SNe Ib/IIb, as does \sn2019cri although this is a few 1000 \kms\ slower than the SNe Ic.
\CaII\ is often found to evolve with similar velocities to \FeII\ \citep{Prentice2019}, this is likely due to both being produced by primordial elements in the ejecta and so both trace the photosphere. This causes some tension for \sn2019unb with it's standard \CaII\ velocity but very low \FeII\ velocity.

\subsubsection{\HeI\ \lam5876}
Finally, a comparison of the \HeI\ measurements for SNe 2019hge and 2019unb shows that they have lower line velocities compared to the bulk of these objects. However, scaling for rise times shows that the observed evolution of both are very similar to the SNe Ib/IIb.
This line is often a blend of \NaI\ and \HeI\ \lam5876 in SNe~Ib, and in Section~\ref{sec:19unb_model} it is shown that modelling can replicate this particular feature well with just \NaI\ D. The strength of the feature pre-peak is comparable to SNe Ib rather than SNe~Ic however, which would suggest that \HeI\ is present at this time.

\subsubsection{Summary}
The line velocities in the photospheric phase of these objects are generally lower than most, if not all of the comparison objects. The velocity curves evolve slowly, which partially reflects the long photometric evolution of these objects. However, the relative line velocities do not scale directly with the relative light curve time scales, for example, although \sn2019unb is approximately 4--5 time broader than a SN Ib its velocities are only approximately half by comparison.
As with normal SNe, the velocities do not increase at any point. However, this is not well constrained as even changes of 10s of percent are below the resolution of the spectra. Slow velocity evolution can be produced by low specific kinetic energies and/or steep gradients in the ejecta density profile.

\begin{figure*}
    \centering
    \includegraphics[scale=0.65]{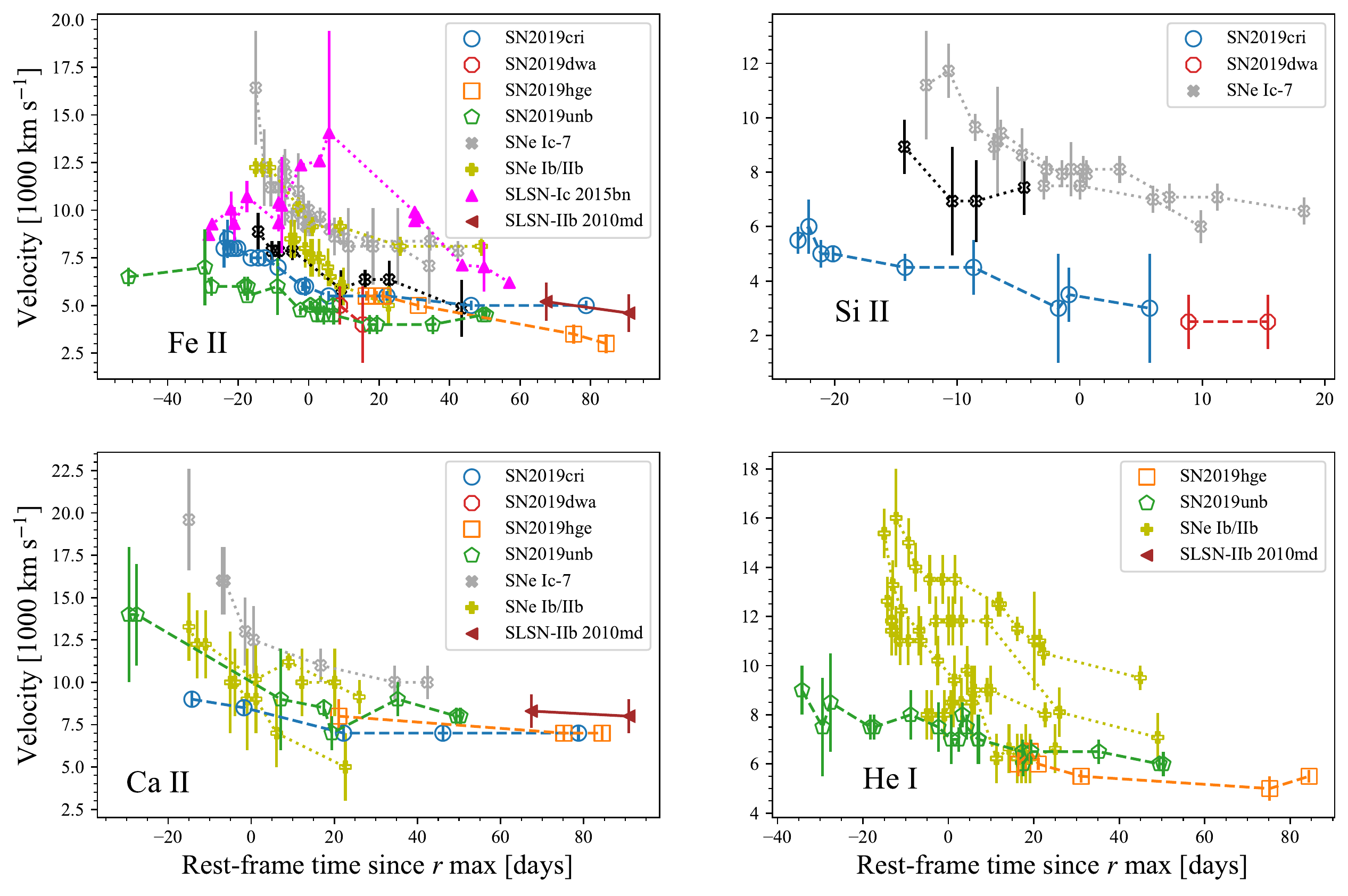}

    \caption{The line velocities measured as a function of time in comparison with SNe Ic-7 (dark grey, SN 2011bm emphasised in black), SNe Ib/IIb (yellow), SLSN-Ic 2015bn \citep[mageneta;][]{Li2017} and SLSN-IIb \citep[brown;][]{inserra2013}. {\bf Top left:} \FeII\ \lam5169. {\bf Top right:} \SiII\ \lam6355. {\bf Lower left:} \CaII. {\bf Lower right:} \HeI\ \lam5876.  
          }
    \label{fig:velocities}
\end{figure*}

\section{Snapshot spectral modelling}
Owing to the similarity of \sn2019cri and \sn2019dwa to SNe Ic, and the lack of obvious H and He in the maximum/pre-maximum spectra of \sn2019unb, we took the opportunity to make a ``snapshot'' model of the maximum light spectrum for each object using a well-established one-dimensional code for the synthesis of SN spectra \citep{Mazzali1993,Lucy1999,Mazzali2000}. The code, based on the Monte Carlo method, uses the Schuster-Schwarzschild approximation and enforces radiative equilibrium in the expanding SN ejecta. It requires as input a density/abundance distribution with radius \citep{2005MNRAS.360.1231S}, as well as an emergent luminosity and the epoch since explosion of the spectrum. Photons emitted at the photosphere are allowed to interact with the gas in the ejecta via absorption processes, which can be followed by re-emission in different lines, thereby implementing both fluorescence and reverse fluorescence processes, and to scatter off free electrons, which increases their residence time in the ejecta. Excitation and ionization are computed using the nebular approximation, which is appropriate in these low density environments \citep{1996A&A...312..525P}.

This code has been applied numerous times on stripped-envelope transients \citep[e.g.,][]{Mazzali2002,Sauer2006,Mazzali2017,Ashall2019,2020MNRAS.497.3542P,Teffs2021}, including SLSNe \citep{Mazzali2016}, and was used for a parameter study in relation to extremely energetic SNe Ic \citep{Ashall2020}.
For our models, we needed to create ad-hoc density profiles. 
Given that the SNe we analysed can be classified as Ic-7, which suggests a similar specific kinetic energy \eom\, but that their light curves indicate a possible range of masses, we started with model CO21 \citep{Nomoto1994}, which was developed to model the spectra and light curves of Ic-6 \sn1994I. The initial abundances were based upon the work of \citet{Sauer2006}.
Next, we set suitable luminosities $L$ and photospheric velocities \vph\ in order to match the flux level and line velocity of the spectrum. 
We rescaled this model in \mej\ and \ek\ using the rescaling equations of of \citet{Hachinger2009}.
We used the assumption that objects with similar spectra have similar \eom\ \citep[e.g.,][]{Mazzali2013,Teffs2021}.
This was then used to constrain \ek\ to \mej, with adjustments made to fit the line widths. The ejecta mass was found by fixing time time of the spectrum $t-t_{\mathrm{exp}}$ and then iterating the mass scaling until viable fits were found with a dilution factor of the model $w$ is between 0.4 and 0.6.
Finally, we adjusted the elemental abundances in order to optimise the fit to the observed spectra.

The process of determining optimal parameters is iterative, but when convergence is achieved it likely to point to the correct values given the model that has been used. While individual parameters can have uncertainties ranging from 5 to 25 per cent, when they are taken all together in a ``best-fit model'', the uncertainty is likely to be much smaller, of the order of 5-10 percent \citep{Ashall2020}.

\subsection{Modelling the maximum light spectrum of SN\,2019cri}\label{sec:19cri_model}
\begin{figure}
    \centering
    \includegraphics[scale=0.6]{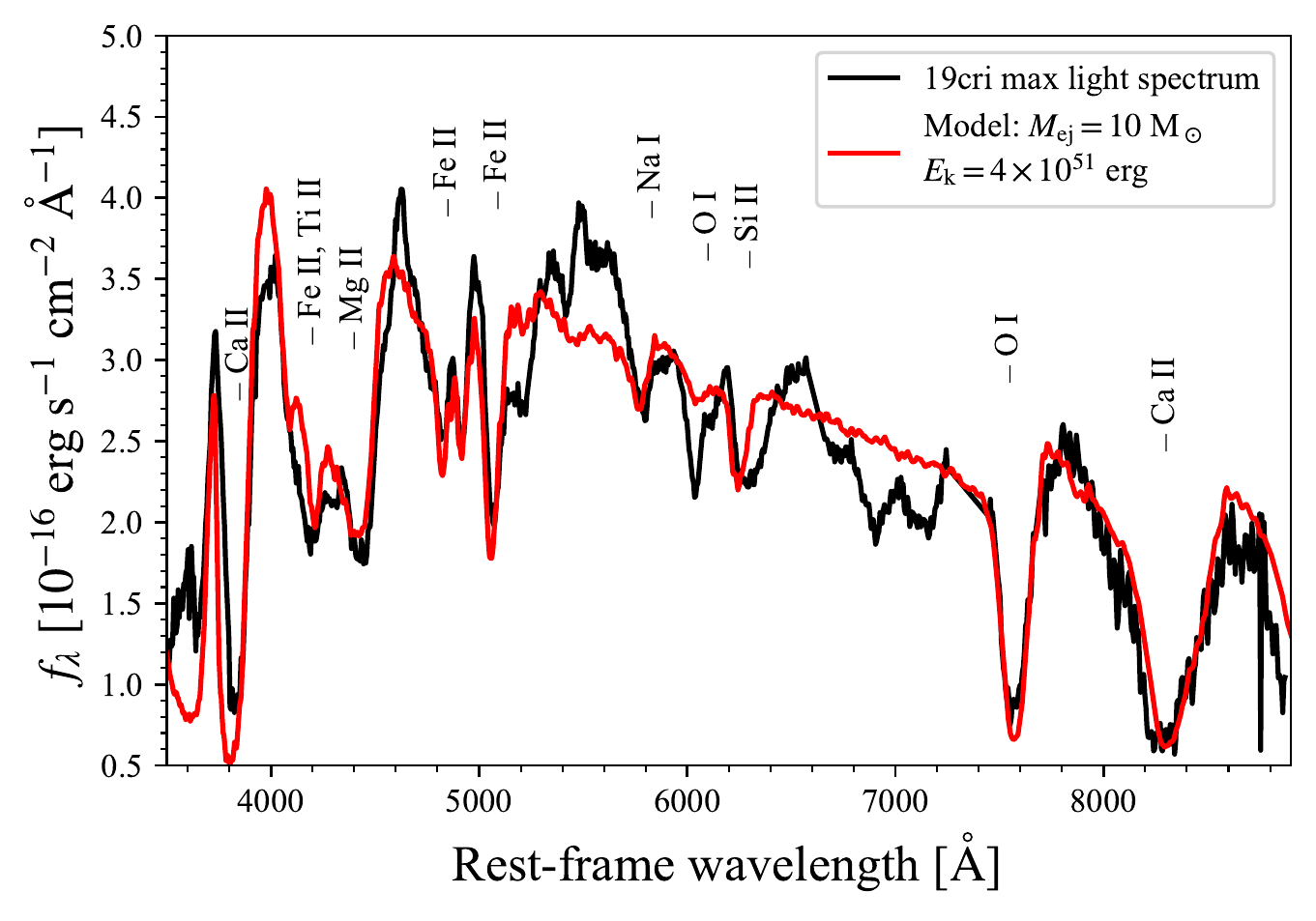}
    \caption{Snapshot modelling of \sn2019cri using a density profile of 10 \msun\ and \ek\ = $4 \times 10^{51}$ erg.
    }
    \label{fig:19cri_model}
\end{figure}

The redshift and extinction corrected maximum light spectrum, in relation to the $r$-band, is dated 2019-05-14 and was observed by the NTT. 
A density profile with \mej\ $= 10$ \msun\ and \ek\ $=4$ foe (foe $=1 \times 10^{51}$ erg), and a specific kinetic energy of 0.4 [foe/\msun] was found to give appropriate model spectra at the required time $t$. 
This compares favourably with the 0.56 [foe/\msun] found for \sn2017ein \citep{Teffs2021}.
The time since explosion $t-t_\mathrm{exp} = 65$ days, with a photospheric velocity \vph\ $=5700$ \kms.
The abundance at this time assuming a single shell is 50\% O, 29\% C, 20\% Ne, with the remaining 1\% split between (in decreasing order) Mg, Si, S, Na, \Nifs, \Fefs, Ti, and Ca.

The kinetic energy was constrained through the width of the lines, in particular the \FeII\ and \MgI\ absorptions around 5000 \AA, which are distinct and are characteristic of SNe Ic-7.
A larger kinetic energy leads to broader lines that blend together. This constraint on the kinetic energy is limited however, as it is sampling the lower velocity material. If the earlier spectra had displayed broader lines then this would have required a shallower density profile at these epochs which results in an increase in kinetic energy without an equivalent increase in ejecta mass, this then would increase the specific kinetic energy \citep{Mazzali2013}.
The photospheric velocity at this epoch is consistent with the measured line velocities in the spectra (see Section~\ref{sec:line_vels}).

Despite this, a point of concern remains; the density profile used to give this snapshot is designed to produce the standard light curve of a supernova and could not replicate the variations seen in the post-peak light curves of \sn2019cri. 
Part of the problem is that energy deposition from the \Nifs\ decay chain decreases in a predictable manner over time, and that the light curve variations are achromatic in the optical.
There are no indicators in the spectra either to suggest this is a recombination effect, as in the secondary bump in the $r$ band light curves of SNe Ia.
Speculatively, this may then suggest that the luminosity variation is due to an increase in continuum flux and may arise from thermalisation of energy emitted from a source in the interior of the ejecta.

\subsection{The $+8$ day spectrum of \sn2019dwa}

\begin{figure}
    \centering
    \includegraphics[scale=0.6]{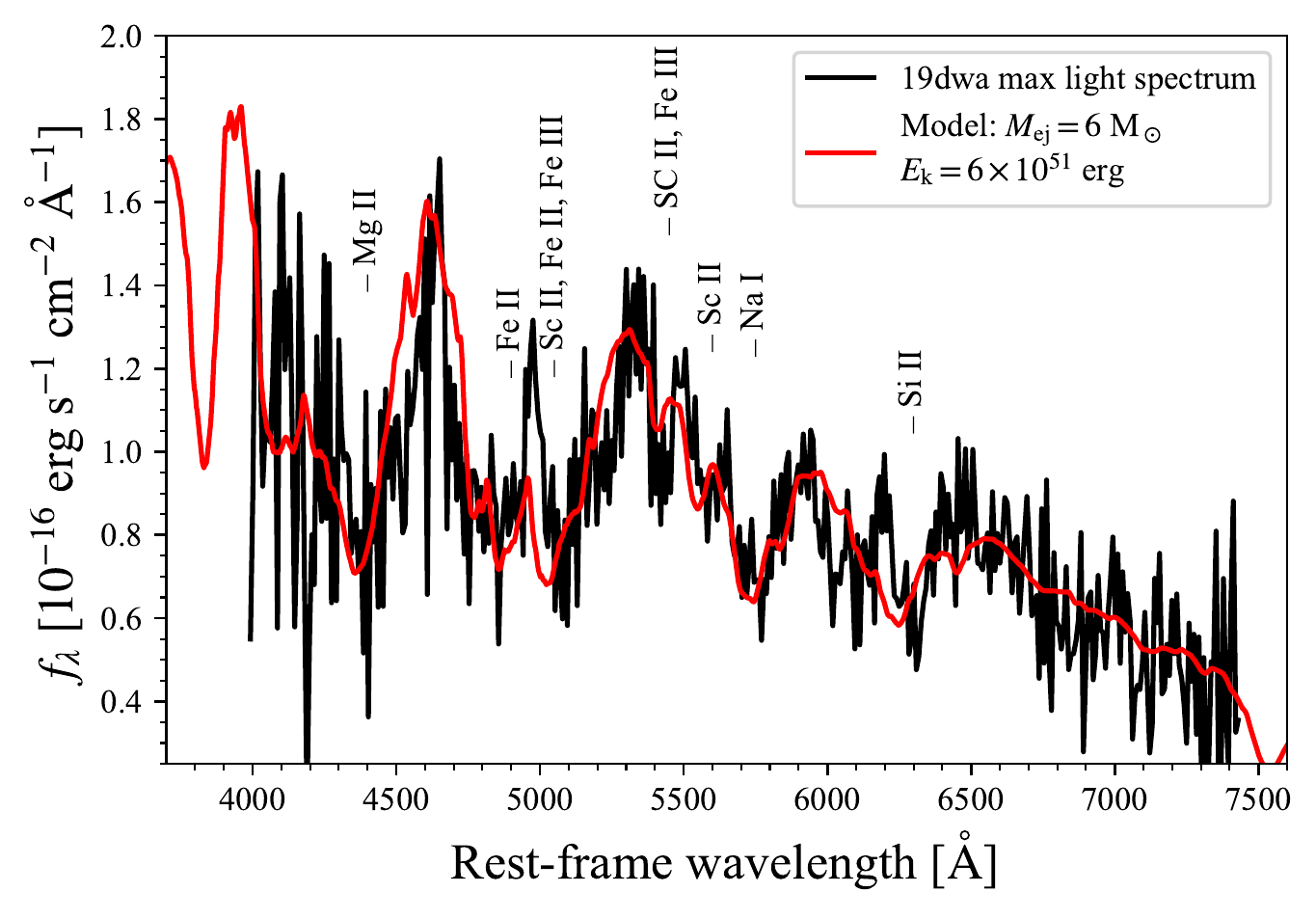}
    \caption{ Modelling of \sn2019dwa using a density profile of 6 \msun\ and \ek\ = $6 \times 10^{51}$ erg.
    }
    \label{fig:19dwa_model}
\end{figure}

For \sn2019dwa, the $+8.7$ day spectrum is used. It has unfortunately low S/N and presents a limited wavelength range of approximately 4000 - 7400 \AA\ in the SN rest-frame.
The spectrum is still quite blue compared with SE-SNe at this relative age, but shows signs of being evolved, such as deep absorption features and low velocities.
A similar process for \sn2019cri was used to find a density profile suitable combination of \mej\ and \ek. 
The provisional model has \mej\ $=6$ \msun\ with \ek\ $=6$ foe, giving \eom\ $=1$ [foe/\msun], at $t-t_\mathrm{exp} = 45$ days with \vph\ $=3000$ \kms.
The ions responsible for the features in the range 4000--5300 \AA\ are mainly \MgII\, \FeII\ and \FeIII. In the intermediate region we find \NaI\ D and \SiII\ \lam6355, with also the likely detection of Sc~{\sc ii} resulting in absorptions, most prominently between 5400--5600 \AA. 
Sc is also identified through spectral modelling in the post-maximum spectra of \sn2017ein \citep{Teffs2021}.
The abundances used in the one-zone model at this time are 25\% O, 20\% Ne, 15\% Si, 15\% C, 10\% Na, 9\% \Nifs, and 1\% Sc and 5\% split between Ca, Mg, \Fefs, and S.

\subsection{SN 2019unb at $+7.2$ days}\label{sec:19unb_model}
\begin{figure}
    \centering
    \includegraphics[scale=0.6]{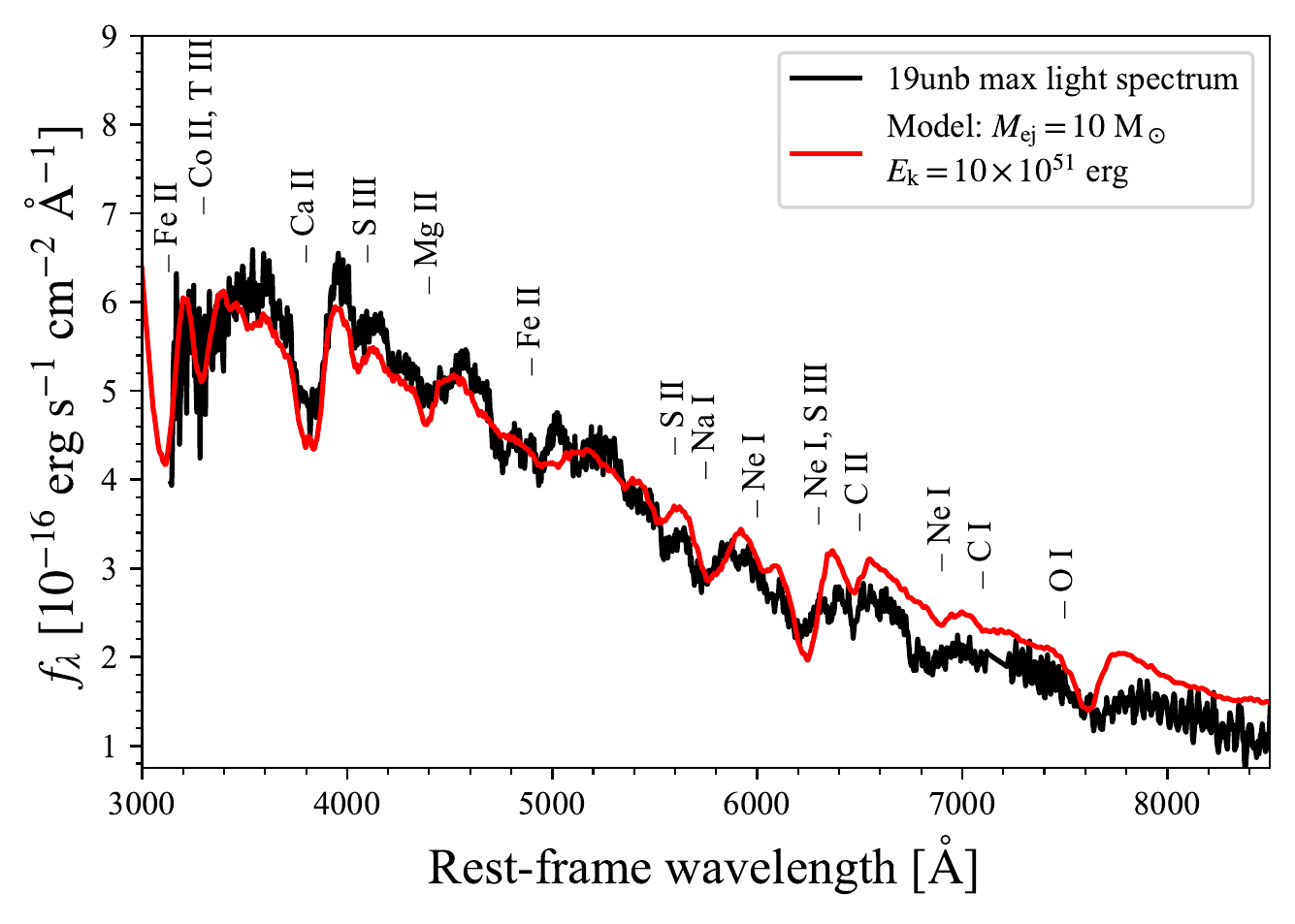}
    \caption{Modelling of \sn2019unb using a density profile of 10 \msun\ and \ek\ = $10 \times 10^{51}$ erg. Time since explosion is 80 days. This snapshot recreates most of the features seen in the observed spectrum. 
    }
    \label{fig:19unb_model}
\end{figure}

The spectrum $+7.2$ days after $r$-band maximum of \sn2019unb is not dominated by H or He, so the same method can be applied to model this spectrum.
Through an iterative process, a density profile with \mej\ $=10$ \msun\ and \ek\ $= 10$ foe was chosen. The time since explosion $t-t_\mathrm{exp} = 80$ days, and photospheric velocity \vph\ $=4900$ \kms.
The low \vph\ supports the \FeII\ velocities that were measured in Section~\ref{sec:line_vels}.
The spectra do not extend blueward enough to constrain the flux below 3200 \AA, but the total bolometric luminosity of the model is $2.6 \times 10 ^{43}$ \ergs ($-19.8$ mag), comparable to the peak of $M_r$ at this time.
The ejecta composition is typical for a massive star; Ca, C, O, S, Si, Fe, Na, Ne. In the model,
the abundances are 35\% O, 40\% C, 20\% Ne, with the remaining 5\% split between Na, Si, S, Mg, \Fefs, and \Nifs.
This leads to an overly strong \OI\ \lam7774 line in the one shell model. A solution to this would be to place some of the excess mass that is in the O fraction into an element that is not strong in the optical, which could be He.
The presence of a ``\Ha\ bump'' in the later spectra would suggest some presence of H, and therefore He, in the ejecta. However, at this phase neither are identified\footnote{Both H and He require NLTE treatment, which is beyond the scope of this work. However, while \HeI\ \lam5876 may contribute some small amount to the \NaI\ feature in the modelled spectrum, H is entirely absent.}.
For the H Balmer features this is likely because the photospheric temperature is $\sim 8000$ K and so H is mostly ionised. 
Helium, on the other hand, needs to be non-thermally excited by the decay products of the \Nifs\ decay chain, but at 80 days most of the \Nifs\ and it's daughter nuclei have decayed to \Fefs.

\section{Discussion}

\subsection{Powering mechanisms}
In addition to the large kinetic energies required, which exceed those for explosion modelling \citep{2015MNRAS.453..287M},  the luminous, long-lived, and variable light curves of these objects leads to the question of the powering mechanism. 
Here we discuss a few potential sources for powering of these events.

The analytical equations of \citet{Arnett1982} can be used to estimate the mass of \Nifs\ required to power the respective light curves. 
This ``back of the envelope'' calculation is used purely to illustrate the difficulty in powering these objects with \Nifs.
If we take the basic approximation that $M_r$ approximated the absolute magnitude at bolometric maximum, then we can convert these values to cgs luminosity units. Combining this with the rise time and using the formulation of ``Arnett's Rule'' given in \citet{Stritzinger2005} we get the following back-of-the-envelope \mni\ estimates of $\sim$1.5 \msun\ for \sn2019cri, $\sim$1 \msun\ for \sn2019dwa, $\sim$3.5 \msun\ for \sn2019hge, and $\sim$4.7 \msun\ for \sn2019unb.
These should be considered an upper limit, as recent explosion models have suggested that ``Arnett's Rule'' could lead to an overestimate of the \Nifs\ mass \citep{2019ApJ...878...56K}.
In each case, the estimated \mni\ would still constitute a considerable fraction of the ejecta mass. This would lead to SN Ia-like temperatures, line blanketing, and emission dominated by Fe-group lines in the later phases, which is not seen. Instead, the results of the spectral modelling suggests that the amount of Fe group elements is relatively comparable to that in normal core-collapse events. This also rules out possibility of these events being pair-instability SNe, as the ejected mass is far too small.

If, as is most likely, that \Nifs\ is not the power source behind these objects the we must look to alternative options. 
An easy way to reach the luminosities seen in these objects along with the variations in the light curves, is to have the ejecta interact with CSM.
In this process, a small fraction of the kinetic energy is converted to radiative energy, and because \ek\ $\sim 10^{51}$ erg for a typical SN, this provides and ample energy reservoir to power the light curves.
Under normal circumstances, the effect of ejecta interacting with CSM is prominent spectroscopic narrow emission lines of H or He, depending on the CSM composition.
This respectively leads to Type IIn and Type Ibn SNe, both of which can reach luminosities greater than the object featured in this work \citep[e.g,][]{Hoss2017,Nyholm2020}.
It was noted in Section~\ref{sec:spectra} that none of these objects show emission lines consistent with CSM interaction.
We can consider this an external process, as the energy injection occurs on the outside of the ejecta, however interaction has been proposed as a way to power the transitional event PTF11rka without explicit spectroscopic signatures of interaction \citep{2020MNRAS.497.3542P}.

An internal process would include radioactive powering (e.g. \Nifs\ and \Cofs\ decay), or some physical process that taps the rotational energy of a compact object.
In this latter category we find black-hole accretions and magnetars.
Magnetars are rapidly rotating highly magnetised neutron stars. 
Their energy reservoir and energy injection rate is governed by two parameters; the period $P$ of rotation and the strength of the magnetic field $B$. 
Magnetars with millisecond spin periods and $B\sim 10^{15}$ Gauss can have a total rotational energy of $>1 \times 10^{52}$ erg and provide a diverse range of monotonically decreasing energy injection rates.
These objects are commonly invoked for SLSNe, as they provide a way to power these objects without requiring tens of solar masses of ejecta and several solar masses of \Nifs, which is difficult to synthesise.
Although most of the spin-down power must be thermalised in order to explain the high continuum luminosities of SLSNe, possible signatures of magnetar powering are present in the spectrum: at early times, broad \OII\ lines seen in most SLSNe and some lower-luminosity SNe could indicate non-thermal ionization in rapidly expanding ejecta \citep[][Parrag et al., submitted]{Mazzali2016}, while at nebular phases a central energy source is needed to power a prominent \OI\ recombination line \citep{Jerkstrand2017,Nicholl2016b,Nicholl2019}. However, the transitional events studied here lack \OII\ lines and we have few nebular phase spectra. Another possible means to test the presence of a central engine is via the evolution of the photospheric velocity.
\citet{Kasen2010} suggested that magnetars may be able to produce cavities at the base of the ejecta equivalent to pulsar winds, with the consequence being that ejecta is pushed from below and accelerated into ejecta above, forming a high density ``wall''. An observational effect of this density spike could be a flat or very slow velocity evolution in spectroscopic lines. 
However, this density spike may also be an artefact of one-dimensional simulations, and can be washed out by turbulent mixing in higher dimensions \citep{Chen2016}, but simulations suggest this is energy-dependent \citep{2019ApJ...880..150S,2021ApJ...908..217S}.

The line velocities in the objects here are indeed slow, amongst the slowest evolving seen in stripped CC-SNe. \sn2019cri provides a good comparison because another explanation for low, and slowly evolving velocities is low energy. However, comparison of the line velocity curves in Fig.~\ref{fig:velocities} with SN~Ic-7 (equivalent specific kinetic energy) shows that this is not an impediment to having a relatively normal velocity profile. 

A central compact object powering these SNe is an attractive idea. The evidence suggests that the powering source is probably not external unless all four objects were able to hide the CSM interaction below their photosphere.
The luminosity variability of such an object, perhaps through accretion, is also a possible way to explain the variations in the light curves without affecting the appearance of the optical spectra. 

\begin{figure}
    \centering
    \includegraphics[scale=0.45]{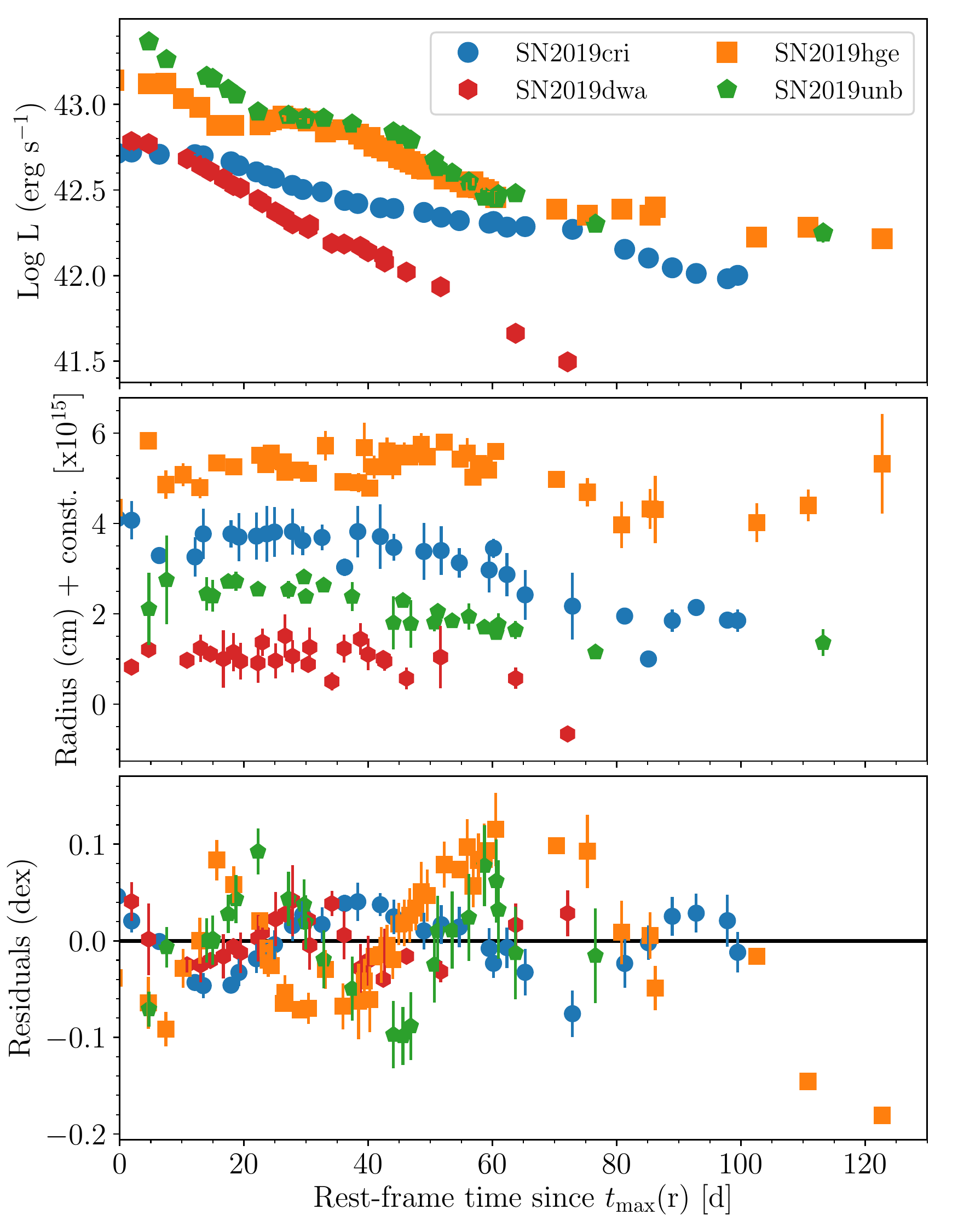}
    \caption{ {\bf Upper panel:} After peak bolometric light curves of our transitional objects. {\bf Middle panel:} photospheric radius evolution as derived from black-body fitting of the spectral energy distribution. {\bf Lower panel:} residuals, after subtracting fits to the declining bolometric light curve.
    }
    \label{fig:undulations}
\end{figure}

\subsection{Undulations and their origin}
As observed in Section~\ref{sec:lcs}, these objects show some degree of undulations in their luminosity evolution. This is reminiscent of what observed in SLSNe~I like SN~2015bn \citep{Nicholl2016,inserra2017}. To explore this further, we fitted a first-order
polynomial to the bolometric light curve from the peak to the end of the available photometry. The choice of using the bolometric allows us to avoid an analysis dominated by line evolution within a particular passband. The bolometric light curves (cfr. Figure~\ref{fig:undulations}, top panel) were built following the procedure outlined in \citet{Inserra2018a}. The choice of a linear fit is due to the fact that SE-SNe are usually well reproduced by models having a steady decline over the available timescale. 
The residuals in Fig.~\ref{fig:undulations} (bottom panel) show clear fluctuation of the order of roughly 0.1 dex for SN~2019hge (at $\sim20$d and $\sim70$d) and SN~2019unb (at $\sim20$d and $\sim60$d) and possibly also in SN~2019cri at $\sim70$d. Nothing conclusive can be said for SN~2019dwa as any possible fluctuation is at the same level of the uncertainties. This reflects what observed in the single broadband filters. Such fluctuations are a factor of 10 larger than the uncertainties, and roughly a factor of 2 larger than what observed in `slow' SLSNe I \citep[e.g.][]{Nicholl2016,inserra2017,inserra2019}.

In a normal SN scenario, such fluctuations might be caused by a change in the density profile of the emitting region. Such a change is also usually reflected by a change in the radius evolution of the photosphere \citep{liu2018}. Hence, we derived the radius evolution via the Stefan-Boltzmann law and using as input the bolometric light curve and the temperature evolution derived by fitting the available spectra energy distribution with a Planck function. The radius evolution displayed in Fig.~\ref{fig:undulations} (middle panel) do not show any stark change in the radius evolution. Nevertheless, a change in the line/ions is observed in the spectra evolution. 
Another viable alternative to produce such undulations is via interaction with a small amount of CSM material as observed in interacting transients \citep[e.g.][]{Smith2012}. Using the scaling relation presented by \citet{SM2007} $L\approx M_{\rm CSM}v^2/2t$ and assuming an average half-period of the fluctuation of $t=10$ day (only the fluctuation at 70d of SN~2019hge has half-period of 20 days), average luminosity of $L=10^{43}$ erg s$^{-1}$ and velocity of 5000 km s$^{-1}$ we obtain CSM masses of $0.03-0.07$ M$_{\odot}$. As this seems periodic, they might be the consequence of a close binary scenario for the progenitor, which cause an heterogeneous structure of the CSM and could also explain the absence of interaction signature in the spectra due to a clumpy structure or viewing angle \citep{Moriya2015}. 

\subsection{\sn2019cri and Ic-7}
\begin{figure}
    \centering
    \includegraphics[scale=0.6]{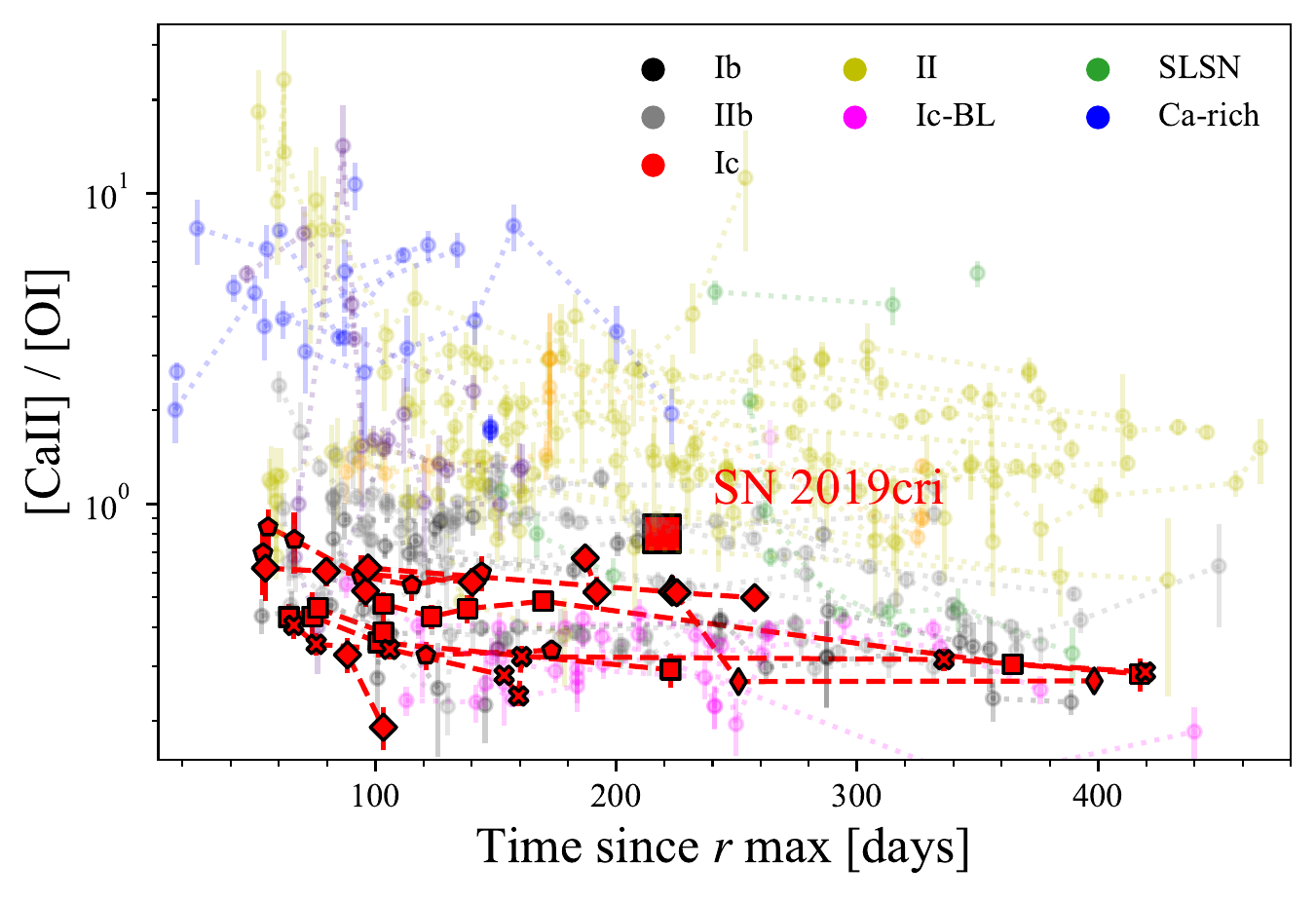}
    \caption{The \caiif/\Oneb\ plane (Prentice et al., submitted) with \sn2019cri (red square) in context against SNe Ic (red) and other transients (see legend). The ratio of \sn2019cri is marginally larger than SNe Ic at the same epoch. It may be expected however, that if this ratio is a measure of core oxygen mass, that \sn2019cri would be lower than the other transients based upon the light curve and snapshot spectral model, which would indicate a relatively large O mass.
    }
    \label{fig:19cri_ratio}
\end{figure}

The narrow lines of SNe Ic-7 are indicative of steep ejecta density profiles which results in low specific kinetic energy. The model spectrum presented here of \sn2019cri agrees with the \eom\ found for \sn2017ein by \citet{Teffs2021}.
It becomes difficult to explain why objects with such diverse photometric properties should display so similar spectroscopic properties.
The possible powering mechanisms for the four SNe featured in this work
were discussed previously, and it was stated that the light curve of \sn2019cri is incompatible with a pure \Nifs\ decay model.
The question then arises; if \sn2019cri need not be powered primarily by \Nifs\ then why should the other objects of this class?
It then becomes pertinent to examine \sn2011bm in this context. This transient requires $\sim0.7$ \msun\ of \Nifs\ to power its peak luminosity, comparable to SNe Ia, and $7-11$ \msun\ of ejecta \citep[via Arnett][]{Valenti2012} to display such a broad light curve. 
Prentice et al. (submitted) show that if the ratio of [\CaII] \lam\lam7292, 7324 to \Oneb\ for this object is compared with other SE-SNe, where the ejecta masses are typically $\sim 3$ \msun, then \sn2011bm takes the same ratio as low mass objects.
In Section~\ref{sec:19cri_spectra} a value of $\sim 0.8$ was calculated for \sn2019cri.
Figure~\ref{fig:19cri_ratio} shows that this objects sits above is an outlier compared with other SNe Ic.
This ratio is assumed to trace oxygen core mass \citep{1987ApJ...322L..15F}, which itself is a proxy for \mej\ in stripped supernovae. 
Prentice et al. (submitted) suggest that one possible reason for this is that the powering mechanism of \sn2011bm is not predominantly via the \Nifs\ decay chain and also pointed to lack of strong \FeII\ and \FeIII\ emission lines in its nebular spectra.
In \sn2019cri we find an object that is more extreme, where \Nifs\ is disfavoured as the primary powering source. Fortunately, this transient was followed into the early nebular phase. By taking the [\CaII]/[\OI] ratio this object can be placed in context, and it is found that the ratio is even larger than for the SE-SNe. This could suggest that the slow evolution of this object is not due to a large ejecta mass, but perhaps a dense wall within the ejecta profile, perhaps blown by a magnetar \citep{Kasen2010}. Another possibility is that the [\CaII] line is contaminated or even dominated by [\OII], artificially inflating the [\CaII]/[\OI] ratio, if oxygen in the ejecta remains ionised into the nebular phase \citep{Jerkstrand2017}. This can occur due to `freeze-in' of the ionisation state in ejecta heated by a power-law central engine \citep{Margalit2018}. 

\section{Conclusions}

\sn2019unb and \sn2019hge have both been classified as superluminous supernovae, yet, photometrically they are close to SNe Ibc in terms of luminosity (they sit close to the $M=-19.8$ mag boundary defined by \citealt{Quimby2018}), strengthening the case that a more robust definition, ideally linked to the spectroscopic evolution, is needed \citep{GalYam2019,inserra2019}.
Spectroscopically they share few similarities with the canonical SLSN, they lack the \OII\ and \CII\ lines seen in their early spectra, but they do not share the same evolution timescales as they are  more similar to SNe Ib/IIb at later times than normal SLSN. It was noted that convolving a black body with a SN Ib spectrum gave a good approximation to the early spectra of \sn2019unb and spectral modelling suggests a 10 \msun\ ejecta and normal relative abundances compared with SN Ib.

SNe 2019cri and 2019dwa, are similar to SE-SNe, especially SN Ic-7; displaying narrow \FeII\ lines. Snapshot spectral modelling of the maximum light spectra suggests these transients may have large ejecta masses or 10 and 6 \msun\ respectively, and a specific kinetic energy of $\sim 1$, although we add a note of caution with these results.
It also shown that \sn2019cri evolves spectroscopically on a timescale typical of SE-SNe after maximum and that the appearance of \Oneb\ coincides with a drop in luminosity.
We suggest that this may be evidence for a density ``wall'' within the ejecta, whereby the line forming region drops from a high to low density region and so the spectra change appearance relatively rapidly. This is analogous to the bubble blown by a pulsar wind and has been shown to occur in the presence of a magnetar.

The variability and luminosity of the light curves preclude powering from a purely radioactive source, while the variability itself suggests either variation in the powering source or a non-monotonic ejecta density profile. Given the absence of any clear emission lines associated with CSM interaction in any of the objects, we suggest that the powering mechanism for the light curve and ejecta dynamics is likely dominated by energy from a compact source.

It is demonstrated that the hosts of these SNe are consistent with that of SLSNe; low mass and high star formation rates, with the exception of \sn2019hge, which was hosted in a high mass galaxy with a high star formation rate. 
Considering the inferred ejecta masses along with the presence of these events in star forming galaxies, it is likely that the progenitors of these objects are more massive than they that typically associated with SE-SNe and more akin to SLSNe.

If these objects are to be placed within the existing taxonomy, the classification should reflect physical parameters of the objects.
In this case we can summarise these two events as follows:

\begin{itemize}
    \item They have He-rich ejecta, possibly also a thin H shell.
    \item They have extremely low ejecta velocities in comparison with other SNe.
    \item Their light curves are luminous and variable which might be a hint of a binary system scenario (for at least some of them).
    \item Their early spectra are hot, and their early colour evolution suggests this is the case for many weeks.
    \item Spectroscopically the display no obvious signs of CSM interaction.
\end{itemize}

We can only hypothesise as to what powers these events, and there exists no theoretical model to explain their observed properties.
Given the spectroscopic similarity with SNe Ib/IIb it may be useful to consider than as a peculiar sub-class of these events.
It would also not be the first time that SNe Ib/IIb have displayed unusual photometric properties. Prior to this was the case of \sn2005bf and its double peaked light curve that reached $M_r \sim -18.5$ in nearly 60 days.

An increasing number of these rare objects are now being found and prove the diversity of SNe between SLSNe~I and SE-SNe. To understand them, we need to address the following questions: do SE-SNe and SLSNe represent a continuous class of objects as suggested by the luminosity function, or are they separable, as suggested by their spectroscopic sequence? Can high resolution spectroscopy show a relationship between the ejecta velocity and light curve variation? If so, this could represent deviation away from a power law ejecta density profile.

\section*{Acknowledgements}
KM, MRM, SJP are supported by H2020 ERC grant no.~758638.
L.G. acknowledges financial support from the European Union's Horizon 2020 research and innovation programme under the Marie Sk\l{}odowska-Curie grant agreement No. 839090, and from the Spanish Ministry of Science, Innovation and Universities (MICIU) under the 2019 Ram\'on y Cajal program RYC2019-027683.
TMB was funded by the CONICYT PFCHA / DOCTORADOBECAS CHILE/2017-72180113.
MG is supported by the EU Horizon 2020 research and innovation programme under grant agreement No 101004719.
SGG acknowledges support by FCT under Project CRISP PTDC/FIS-AST-31546/2017.
MN is supported by a Royal Astronomical Society Research Fellowship and H2020 ERC grant no.~948381.
T.-W.C. acknowledges the EU Funding under Marie Sk\l{}odowska-Curie grant H2020-MSCA-IF-2018-842471
The Liverpool Telescope is operated on the island of La Palma by Liverpool John Moores University in the Spanish Observatorio del Roque de los Muchachos of the Instituto de Astrofisica de Canarias with financial support from the UK Science and Technology Facilities Council.
Based on observations collected at the European Organisation for Astronomical Research in the Southern Hemisphere, Chile, as part of ePESSTO+ (the advanced Public ESO Spectroscopic Survey for Transient Objects Survey). ePESSTO+ observations were obtained under ESO program ID 1103.D-0328 (PI: Inserra).
The WHT is operated on the island of La Palma by the Isaac Newton Group of Telescopes in the Spanish Observatorio del Roque de los Muchachos of the Instituto de Astrofísica de Canarias.
SJP thanks GPL for many insightful discussions at the bar over the last few years.

\section*{Data Availability}
Data will be made available on the Weizmann Interactive Supernova Data Repository (WISeREP) at www.wiserep.org.




\bibliographystyle{mnras}
\bibliography{bib.bib} 




\appendix

\section{SN 2019yz}
\sn2019yz is a spectroscopically and photometrically normal Type Ic-7, and is provided here as a comparison object. We present a quick summary of the properties of this transient in Table~\ref{tab:19yz}. 
The distance modulus is a Tully-Fisher measurement, while \Emw\ is from \citet{Schlafly2011}. The host extinction is estimated by fitting two Gaussian profiles to the host \NaI\ D absorption lines and applying the relationship between the pseudo-equivalent width and reddening from \citet{Poznanski2012}.
The data for this object will be made public along with the SNe featured in this work.

\begin{table}
    \centering
    \caption{The properties of SN 2019yz}
    \begin{tabular}{ccccc}
        SN & $z$ & $\mu$ &  \Emw\ & \Eh  \\
           &    &  [mag] & [mag] & [mag] \\ 
    \hline       
        2019yz & 0.006 & 32.3 & 0.1 & 0.2$\pm{0.1}$ \\
    \hline    
    \end{tabular}
    
    \label{tab:19yz}
\end{table}



\bsp	
\label{lastpage}
\end{document}